\documentclass{article}
\usepackage{amsfonts}
\usepackage{amssymb}
\usepackage{graphicx}
\usepackage{epstopdf}
\usepackage{cite}
\usepackage[usenames]{color}
\usepackage{graphicx}
\usepackage{epstopdf}

\pdfoutput=1
\providecommand{\U}[1]{\protect\rule{.1in}{.1in}}
\newcommand{\ba}{\begin{array}}
\newcommand{\ea}{\end{array}}
\newcommand{\Dsl}[1] { \setbox0=\hbox{$#1$}     
\dimen0=\wd0   \setbox1=\hbox{/} \dimen1=\wd1  \ifdim\dimen0>\dimen1        
 \rlap{\hbox to \dimen0{\hfil/\hfil}}  #1 \else \rlap{\hbox to \dimen1{\hfil$#1$\hfil}}  /  \fi  }
\newcommand{\bea}{\begin{eqnarray}}
\newcommand{\eea}{\end{eqnarray}}

\input{tcilatex}
\begin{document}

\title{
 {\Large  Exclusive  decays $\chi_{cJ}\rightarrow K^*(892)K $ within the
effective field theory framework} 
}
\author{ Nikolay Kivel  \\[2mm]
\textit{Institut f\"ur Kernphysik, Johannes Gutenberg-Universit\"at, D-55099, Mainz, Germany } 
\\
\textit{and}
\\
\textit{Petersburg Nuclear Physics Institute, Gatchina, 188300, St.~Petersburg, Russia}  
}

\maketitle

\begin{abstract}
We study  hadronic decays $\chi_{cJ}\rightarrow K^*(892)\bar{K} $ within the effective field theory framework.  
We consider  the colour-singlet and  colour-octet contributions and study their properties using  (p)NRQCD effective theory.   
 We show that  infrared singularities in collinear integrals of the colour-singlet amplitudes  can be absorbed into the renormalisation of  the colour-octet matrix elements.  The heavy quark spin symmetry allows us  to establish a relation between  the colour-octet matrix elements and to define the spin symmetry breaking corrections which are free from infrared singularities.  We apply obtained results for a phenomenological description of the  branching fractions. 
\end{abstract}

\noindent

\newpage

\section{Introduction}
\label{int}

A study  of  heavy quark systems like charmonium and bottomonium has been one of the most interesting topic of particle physics  for many years already. 
Many new interesting experimental results have been obtained by  BABAR, BELLE, BESII and BESIII collaborations during  last  years.  
In particular, many new data about various exclusive decays  have been collected and many new results  are expected in the future.   
On the other side a theoretical description  of  various exclusive decay channels  remains puzzling,  see e.g. discussions in  reviews  \cite{Brambilla:2004wf, Brambilla:2010cs} and references therein.  Often,  underlying hadronic  dynamics is very complicated and  involves  non-perturbative effects  which are even difficult to include into a systematic  theoretical description.  One of  such problematic contributions  is  the colour-octet mechanism \cite{Bodwin:1992ye, Brambilla:2004wf}.   In inclusive processes such contributions are described as unknown long-distance matrix elements \cite{Bodwin:1994jh}  but  for  exclusive decays a systematic description of  such mechanism is still not well understood \cite{Brambilla:2004wf, Brambilla:2010cs}.  At the same time such contributions can play an important role in the correct  description of various exclusive amplitudes.   In Ref.\cite{Chen:1998ma}  it is suggested  that the colour-octet  configuration may play an important role for an understanding of the well known ''$\rho\pi$-puzzle``.   Some attempts  to build a framework for  description of  the colour-octet matrix elements  can be found in Refs.\cite{Bolz:1996wh, Bolz:1997ez,Wong:1999hc}.   In Ref.\cite{Beneke:2008pi} it was shown that  a correct description of   colour-singlet  amplitudes with infrared divergencies  is related to the contribution of the   colour-octet  matrix elements. 

 In the present  work  we consider hadronic  decays $\chi_{cJ}\rightarrow K^\ast \bar{K}$  which are  interesting because of specific properties of the corresponding amplitudes.  The  branching fractions of these decays  have been  measured by the BES collaboration  \cite{Ablikim:2006vm,BESIII:2016dda} . In Table~\ref{dataXcJ}   we collect experimental results from  \cite{Patrignani:2016xqp}.
 \begin{table}[th]
\centering
\begin{tabular}{|c|c|c|}
\hline
 $\chi_{cJ}\rightarrow VP $ & $ K^{\ast}(892)^0\, \bar K^0$+c.c.  &  $ K^\ast(892)^+\,\bar K^-$+c.c.   \\ \hline
$\chi_{c1}$& $10\pm 4$  & $15\pm 7$  \\ \hline
$\chi_{c2} $ & $1.3\pm 0.28 $ & $1.5\pm 0.22$   \\ \hline
\end{tabular}%
\caption{ The branching fractions  $\chi _{cJ}\rightarrow K^\ast K $  in units of $10^{-4}$. 
}
\label{dataXcJ}
\end{table}

The  amplitudes for these  decays  are closely  related to  the $SU(3)$  flavour  symmetry  breaking  effects in QCD  and the experimental results for the decay rates  indicate  that such  contributions  are sufficiently  large. 

 Another interesting  point  is that the decay amplitude of  tensor state $\chi_{c2}$  is suppressed according to the helicity selection rule  \cite{Brodsky:1981kj, Chernyak:1981zz, Chernyak:1983ej}.   Hence,  this  amplitude is  sensitive to higher Fock components of  mesonic wave functions.  A sufficiently large value of the measured decay rates implies  the strong violation of the helicity selection rule.  In this respect this process could be similar to the decay $J/\psi\to \rho\pi$ and probably  have  resembling underlying decay mechanism. 
  
In Ref.\cite{Liu:2009vv}  it is suggested that the amplitude for $\chi _{c2}\to K^\ast K$ decay is dominated by a long distance decay mechanism which can be accounted through a model with intermediate mesonic loops. The obtained numerical estimate is  about a factor two smaller then the experimental result.  The second decay   $\chi _{c1}\to K^\ast K$   has not yet  been discussed in the literature and  we could not find any theoretical predictions for the corresponding decay width.   

In our work we  consider both decays  within  the effective field theory framework.    We apply  NRQCD \cite{Lepage:1992tx, Bodwin:1994jh} and  potential NRQCD (pNRQCD) \cite{Pineda:1997bj,Pineda:1997ie,Beneke:1997zp,Brambilla:1999qa,Brambilla:1999xf,Brambilla:2004jw} effective theories  and soft collinear effective theory (SCET) \cite{Bauer:2000ew, Bauer2000, Bauer:2001ct,Bauer2001,Beneke:2002ph,Beneke:2002ni}   in order to  describe  decays of $P$-wave quarkonia  into $K^*\bar K$  mesons.   An advantage of this framework is  the opportunity  to apply  the  heavy quark spin symmetry (HQSS)  which allows one  to constrain  a  contribution associated with the  colour-octet mechanism.  The latter can play an important role in the understanding of underlying  mechanism of  $P$-wave quarkonia decays \cite{Brambilla:2004wf, Bolz:1996wh, Bolz:1997ez}.     

The computation of  colour-singlet contributions in the helicity suppressed decays  involves  different  twist-2 and twist-3 $K$-meson light-cone distribution amplitudes (DAs).  However such contributions often  have  infrared (IR) divergencies that appear  in the collinear convolution integrals.  Then,  a naive collinear factorisation is violated and colour-singlet mechanism  can not be considered as only  one possible contribution.  Such  situation often arises in the description of amplitudes  involving  the higher Fock components of  hadronic wave functions. Rigorously speaking,  a systematic  description of  such endpoint divergencies still remains challenging to theory. 

 Sometimes the structure of  the IR divergencies allows one  to conclude about the presence of  a colour-octet matrix element. Such situation has been  considered  for amplitudes of  $B\to \chi_{cJ}K$ decays in Ref.\cite{Beneke:2008pi}.  In this work it is shown that the endpoint singularities in the colour-singlet contribution  can be absorbed into a colour-octet operator matrix element computed in the  Coulomb limit.  
 
 In the present work we use the same idea,  we will define and compute the relevant  colour-octet matrix elements in the Coulomb limit.  We also study  the  HQSS  constraints for  the colour-octet matrix elements of $P$-wave charmonia which  are dictated  by the structure of the effective Lagrangian where the interactions with the  heavy quark spin are suppressed.  The existence of a  relation between  the colour-octet matrix elements  allows one  to  define  a consistent  IR subtraction  scheme  for a calculation of  the spin symmetry breaking terms. Such technique, also known as  physical subtraction scheme,  is successfully used for the description of various amplitudes in $B$-decays, see {\it e.g.} Refs.\cite{Beneke:2000wa,Beneke:2001ev}. 

Our paper is organised as follows: in Sec.~2 we set up the notation and  define kinematics and amplitudes.  In Sec.~3 we compute  various  colour-singlet contributions and study their properties.  Sec.~4  is devoted to  analysis of the colour-octet contributions in the Coulomb limit.  The Sec.~5  is devoted to a phenomenological consideration.  We  discuss  effects provided by  the   symmetry-breaking  corrections  and estimate a contribution of  the colour-octet matrix elements. Then we conclude in Sec.~6.

\section{Kinematics, notation  and decay amplitudes}
The decay amplitudes  $\chi
_{cJ}\to \bar{K}+K^\ast$ are defined as%
\begin{equation}
\left\langle \bar{K}(k)K^\ast(p);\text{out}\right\vert \left. \text{in};\chi
_{cJ}(P)\right\rangle =i(2\pi )^{4}\delta (P-p-k)\ \mathcal{M}_{\chi
_{cJ}\rightarrow \bar{K}K^* }.
\end{equation}%
 In what follows we use the frame where  heavy meson is at rest and the $z$-axis is
chosen along the momenta of  outgoing particles
\begin{equation}
P=M(1,\vec{0})=M\omega ,  \label{def:w}
\end{equation}%
where $M$ is charmonium mass  and $\omega $ denotes charmonium  four-velocity. 
Any four-vector $V$  which is orthogonal  to velocity $\omega$ is denoted with the subscript  $\top$: $\omega \cdot V_{\top} =0$. 

The  momenta of the outgoing mesons read
\begin{equation}
k=(k_0, 0,0, k_z),~ p=(p_0,0,0, p_z),
\end{equation}%
with (for simplicity, in the following  we use $m_{\bar K}\equiv m_P$ and $m_{K^*}\equiv m_V$ )
\begin{align}
k_{0}&=\frac{M^{2}+m_{P}^{2}-m_{V}^{2}}{2M},~\ p_{0}=\frac{M^{2}-m_{P}^{2}+m_{V}^{2}}{2M},\,
\\
k_{z}&=-p_{z}=\frac{1}{2M}\left[  \left(  M^{2}-\left(  m_{V}-m_{P}\right)
^{2}\right)  \left(  M^{2}-\left(  m_{V}+m_{P}\right)  ^{2}\right)  \right]
^{1/2}.
\end{align}
Assuming  that the heavy quark mass is sufficiently  large $m_c\gg \Lambda _{QCD}$ one obtains
\begin{equation}
~k\simeq m_c(1,0,0,1)=2m_c\frac{\bar{n}}{2},~\ \ p\simeq m_c(1,0,0,-1)=2m_c\frac{n}{2},
\label{def:nnb}
\end{equation}%
where we introduced auxiliary light-cone vectors $n$ and $\bar{n}$ with $(n%
\bar{n})=2$.  Any four-vector $V^{\mu }$  can be decomposed  as   
\begin{equation}
V^{\mu }=\left( V\cdot n\right) \frac{\bar{n}^{\mu }}{2}+\left( V\cdot \bar{n%
}\right) \frac{n^{\mu }}{2}+V_{\bot }^{\mu },
\end{equation}%
where $V_{\bot }$ denotes the components which are  transverse to the light-like
vectors : $\left( V_{\bot }\cdot n\right) =\left( V_{\bot }\cdot \bar{n}%
\right) =0$.  In particular, in the rest frame\begin{equation}
\omega =\frac{1}{2}\left( n+\bar{n}\right) ,~\ \omega ^{2}=1.
\end{equation}%
In the following we also use  short  notations
\begin{align}
g_\perp^{\mu\nu}=g^{\mu\nu}-\frac12(n^\mu\bar{n}^\nu+n^\nu\bar{n}^\mu), \, \,  i\varepsilon^\perp_{\mu\nu}= \frac12 i\varepsilon^\perp_{\mu\nu\alpha\beta}n^\alpha \bar{n}^\beta,
\end{align}
with   $\varepsilon_{0123}=1$.
The decay amplitudes  can be parametrised as
\begin{align}
\mathcal{M}_{\chi_{c1}\to \bar{K}K^* }&  =\left( \epsilon_{\chi} \cdot k\right)  \left(
e_{V}^{*}\cdot k\right)  \frac{m_{V}}{M^{2}}\mathcal{A}_{1}^{\Vert}+\left(  \epsilon_{\chi\bot}\cdot e_{V\bot}^{*}\right)  \frac{(kP)}%
{M}\mathcal{A}_{1}^{\bot},~~\label{M1:def}\\
\ \ \mathcal{M}_{\chi
_{c2}\rightarrow \bar{K}K^* }&  =\epsilon_{\chi}^{\mu\nu}k_{\nu} i\varepsilon
_{\mu\alpha \beta\rho}(e_{V}^*)^{\alpha} \frac{k^{\beta}p^\rho}{(kp)}~\mathcal{A}^{\bot}_{2}, \label{M2:def}%
\end{align}%
where $\epsilon_{\chi}$ and $e_{V}^{*}$  denote polarisation vectors of the charmonium states and  vector meson, respectively.  The polarisation vectors satisfy
\begin{align}
\sum_\lambda (\epsilon_{\chi}^{(\lambda)})_{ \mu}(\epsilon_{\chi }^{(\lambda)})^\ast_{\nu}&=-g_{\mu\nu} + P_\mu P_\nu/M^2,
\\
\sum_\lambda (\epsilon_{\chi}^{(\lambda)})_{ \mu\nu}(\epsilon_{\chi }^{(\lambda)})^\ast_{\rho\sigma}
  &= \frac12 G_{\mu\rho} G_{\nu\sigma}+\frac12 G_{\mu\sigma} G_{\nu\rho}  -\frac13 G_{\mu\nu} G_{\rho\sigma}\,, 
\end{align}
where $G_{\mu\nu} = g_{\mu\nu} - P_\mu P_\nu/M^2$,  
the normalization is such that $(\epsilon_{\chi}^{(\lambda)})_{\mu\nu}(\epsilon_{\chi}^{(\lambda')})^\ast_{\mu\nu}= \delta_{\lambda\lambda'}$,  $(\epsilon_{\chi}^{(\lambda)})_{\mu}(\epsilon_{\chi}^{(\lambda')})^\ast_{\mu}= \delta_{\lambda\lambda'}$ and similar for the $K^*$ vector meson. 

The  kinematical factors in Eqs.(\ref{M1:def}) and (\ref{M2:def})  are chosen  in order  to have dimensionless amplitudes $\mathcal{A}_{i}$.  One can easily see that amplitude $\mathcal{A}_{1}^{\Vert}$ describes  decay of the longitudinally polarised $\chi_{c1}$   while the amplitudes  $\mathcal{A}^{\bot}_{1,2}$ correspond to   transversely polarised  $\chi_{c1,2}$.  The expressions for decay widths read
\begin{equation}
\Gamma\lbrack\chi_{c1}\rightarrow \bar{K}K^*]=\frac{|\vec{k}|}{8\pi}\frac{2}{3}%
\frac{k_{0}^{2}}{M^{2}}\left(  \left\vert \mathcal{A}_{1}^{\bot}\right\vert
^{2}+\frac{1}{2}\frac{(pk)^{2}}{M^{4}}\left\vert \mathcal{A}_{1}^{\Vert
}\right\vert ^{2}\right)  ,
\label{Gamma1}
\end{equation}%
\begin{equation}
\Gamma\lbrack\chi_{c2}\rightarrow \bar{K}K^*]=\frac{|\vec{k}|}{8\pi}\frac{1}{5}%
\frac{k_{0}^{2}}{M^{2}}\left\vert \mathcal{A}^\bot_{2}\right\vert ^{2}\left(
1-\frac{m_{P}^{2}}{k_{0}^{2}}\right)  \left(  1-\frac{m_{P}^{2}m_{V}^{2}%
}{(kp)^{2}}\right) .
\label{Gamma2}
\end{equation}

\section{Colour-singlet contributions }
\label{sing}
\subsection{Colour-singlet contribution to amplitude $\mathcal{A}_{1}^{\Vert}$}
\label{col-sing-long}
\begin{figure}[ptb]%
\centering
\includegraphics[
width=4.0in
]%
{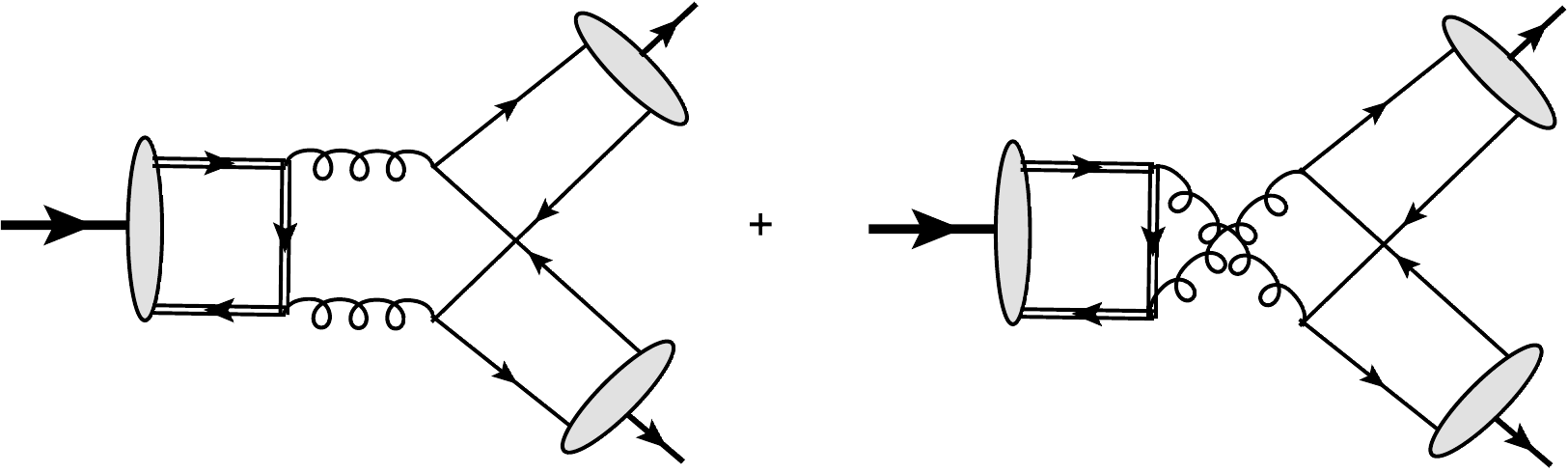}%
\caption{The QCD diagrams describing the colour-singlet mechanism of $\chi_{cJ}\to VP$ decays. The blobs denote various  non-perturbative matrix elements. }%
\label{diagrams}%
\end{figure}

A computation of the colour-singlet contribution is quite standard, corresponding contribution is described by the diagrams in Fig.\ref{diagrams}.   The heavy quark and antiquark  annihilate at  short distance of order $1/m_c$  into the two highly virtual gluons which further create  light quark-antiquark pairs forming the final mesons.  An average  size of  the charmonium  is of order  $1/m_cv$ where $v$ is the  heavy quark velocity in the rest frame.  Since $m_cv\ll m_c$  the colour-singlet decay amplitude is proportional to the heavy meson wave function at the origin.  Corresponding  contribution can be described  by a matrix element of the  appropriate colour-singlet  operator in NRQCD framework.

  Transitions of the light quarks into final mesons also involve non-perturbatibe QCD interactions associated with the typical hadronic scale $\Lambda\ll m_c$.  In charmonium rest frame  energies of the outgoing mesons are large, of order $m_c$  and corresponding  non-perturbative contributions are described by the  light-cone matrix elements  which are related to the light-cone wave functions  at zero transverse separation, the so-called  light-cone distribution amplitudes (DAs).  Detailed description  of these quantities is  given in Appendix~A.  
  
  All  matrix elements arising in  description of the amplitudes  can be estimated according to the power counting with respect to small parameters: velocity $v$ and   ratio $\Lambda/m_c$.  At the leading-order we only have  contribution  to  amplitude $\mathcal{A}^\Vert_1$.  In this case the  soft overlaps with the $in$ and $out$  mesonic sates are described by the leading-order NRQCD matrix element   and by  the leading twist DAs $\phi_{2V}^{\Vert}$ and $\phi_{2P}$ where the subscripts $V$ and $P$ denote the vector and pseudoscalar mesons.  In the following  we  always assume $V\equiv K^\ast$ and $P\equiv \bar{K}$.  The  transverse amplitudes $\mathcal{A}^\perp_{1,2}$ are suppressed by the  power of $\Lambda/m_c$ due to the helicity conservation in the hard subprocess.  As a result  they depend on the twist-3  DAs and this provides suppression by  extra power of the small  ratio  $\Lambda/m_c$.

   The computation  of the diagrams in Fig.\ref{diagrams} with the appropriate operator projections gives the following result
  \begin{equation}
\mathcal{A}_{1}^{\Vert}=-\frac{~  f_{V}^{\Vert} f_{P}}{m_c^{2}}\frac
{i\left\langle \mathcal{O}(^{3}P_{1})\right\rangle }{m_c^{3}}\left(  \frac
{\pi\alpha_{s}(\mu^2)}{N_{c}}\right)  ^{2}~C_{F}~J_{c}^{\Vert}(\mu),
\label{A1Vert}
\end{equation}
with the collinear convolution integral ($1-x\equiv \bar{x}$)
\begin{equation}
J_{c}^{\Vert}(\mu)=\int_{0}^{1}dx~\frac{\phi_{2V}^{\Vert}(x,\mu)}{x\bar{x}}\int_{0}%
^{1}dy~\frac{\phi_{2P}(y,\mu)}{y\bar{y}}\frac{y-x}{xy+\bar{x}\bar{y}}.
\label{def:Jc}
\end{equation}
We also use the  standard notation $C_F=(N_c^2-1)/(2N_c)$ with  $N_c=3$.  The factorisation scale $\mu$ is of order of the hard scale $m_c$.   The definitions   of the non-perturbative constants $f_{V}^{\Vert},\,  f_{P}$ and  $\left\langle \mathcal{O}(^{3}P_{1})\right\rangle$ can be found in Appendix~A.   According to NRQCD counting rules  $\left\langle \mathcal{O}(^{3}P_{1})\right\rangle\sim v^4 $  and ratio $f_{V}^{\Vert} f_{P}/m_c^2\sim (\Lambda/m_c)^2$. Hence  from Eq.(\ref{A1Vert}) one obtains
 \begin{equation}
\mathcal{A}_{1}^{\Vert}\sim v^4 \left(\frac{\Lambda}{m_c}\right)^2.
\end{equation}
From Eq.(\ref{def:Jc})  one can see that  the hard kernel  is  antisymmetric with respect to interchange $\{x,y\}\to \{\bar{x},\bar{ y}\}$  and therefore  the collinear integral is proportional to  antisymmetric  combinations $\phi_{2V}^{\Vert}(x)-\phi_{2V}^{\Vert}(\bar{x})$ or $\phi_{2P}(y)-\phi_{2P}(\bar{y})$  in Eq.(\ref{def:Jc}). Such  combinations do not vanish  for $K$-meson DAs  due to the $SU(3)$ breaking. Using  models for the distribution amplitudes as in Eqs.(\ref{mod:phiP}) and (\ref{mod:phiIIV})  one obtains
\begin{equation}
J_{c}^{\Vert}=\frac{27}{2}\left(  \pi^{2}-4\right)  \left(  b_{1}(\mu)-a_{1V}^{\Vert}(\mu)\right)  +\frac{27}{2}\left(  6\pi^{2}-44\right)  \left(
b_{1}(\mu)a_{2V}^{\Vert}(\mu)-a_{1V}^{\Vert}(\mu)b_{2}(\mu)\right), 
\end{equation}
where  $b_{iP}$ and $a_{iV}^{\Vert}$ are  parameters of the DAs,  see  Appendix~A.  The moments $a^{\Vert}_{1V}$ and $b_1$ vanish in the exact $SU(3)$ limit  which explicitly demonstrates the  dependence of the integral $J_{c}^{\Vert}$ from the flavour symmetry violation.       

Consider  the branching fraction of $\chi_{c1}$ state  assuming that  the transverse amplitude $\mathcal{A}_{1}^{\perp}$ is  small and can be neglected.  In order to obtain  numerical estimate we take  $c$-quark mass  $m_c=1.5$~GeV$,\Lambda^{(4)}_{QCD}=310$~MeV (this gives $\alpha_s(2m^2_c)=0.29$),   the total width  $\Gamma[\chi_{c1}]=0.84$~MeV.  Numerical values of other  parameters are given in Appendix~A. Varying  the factorisation scale $\mu^2$ between $m_c^2$ and $4m_c^2$  we obtain 
\begin{equation}
\text{Br}[\chi_{c1}\rightarrow \bar{K^0}K^*(892)^0+c.c.]= \left(  0.02-0.06\right)\times10^{-3}.
\end{equation}  
We see that this value is about two orders of magnitude smaller then the experimental  branching fraction, see  Table \ref{dataXcJ}.  This result allows one to conclude that the dominant  numerical contribution is most probably   provided by the  amplitude  $\mathcal{A}_{1}^{\perp}$.  This conclusion does also agree with  sufficiently large value of the branching ratio for the $\chi_{c2}$ decay.   

\subsection{Colour-singlet contributions to amplitudes $\mathcal{A}_{1,2}^{\bot}$}
\label{col-sing-perp}

Calculation of the colour-singlet contributions to amplitudes  $\mathcal{A}_{1,2}^{\bot}$  is  more complicated because there are two different configurations: twist-2 and twist-3 projections for $\bar{K}$ and   $\bar{K}^{*}$ states, respectively  ($P_2 V_3$ contribution ) and vice versa ($P_3 V_2$ contribution).  In general, the twist-3 projections  include  contributions from  two-particle and three-particle operators.  The  matrix elements of  three-particle  operators are given by the quark-gluon operators which are  often referred as  genuine twist-3  contributions.  Using QCD equation of motions  the matrix elements of   two-particle twist-3 operators can be rewritten  in terms of twist-2  and  genuine twist-3  quark-gluon DAs, see {\it e.g.} Ref.\cite{Ball:2006wn}. In this work we neglect  the contributions of three-particle quark-gluon operators  in order to simplify  our analysis.  Discarding of the genuine twist-3 contributions is not a rigorous approximation but in a phenomenological calculations  it is often considered as a reliable estimate of higher twist effects.\footnote{Let us also add that such approximation does not contradict to the Lorentz and gauge symmetries in QCD}     Hence we need to consider only the matrix elements of   two-particle  twist-3  operators neglecting the quark-gluon DAs. Such approximation is also known  as Wandzura-Wilczek (WW) approximation. In this case one has to compute the same diagrams as in Fig.\ref{diagrams}  but keeping only twist-2 DAs in  the  twist-3  projections for two-particle  collinear matrix elements.  In order to make our notations simpler  we do not introduce any special notation for  twist-3 DAs in the WW approximation assuming that  this  is clear from the context.     

The calculation is quite standard and we do not discuss here the technical details.  
 The following results has been obtained( remind  that $V\equiv K^{*},\, P\equiv \bar{K}$)
\begin{equation}
\mathcal{A}_{Jc}^{\bot(0)}=\frac{i\left\langle \mathcal{O}(^{3}P_{J})\right\rangle }{m_c^{3}}\left(\frac{\pi\alpha_{s}}{N_{c}}\right)  ^{2}C_{F}\, 2^{J/2}
\left\{
\frac{f_{P}f^{\Vert}_{V}m_{V} }{m_c^{3}}  J_{c}^{(J)}[P_{2}V_{3}]+\frac{f_{P}\mu_{P}f_{V}^{\bot} }{m_c^{3}} J_{c}^{(J)}[P_{3}V_{2}]
\right\},
\label{res:A0Jbot}
\end{equation}
where  the subscript ``$c$" is introduced in order to stress the collinear operator structure for the final mesonic state.  
The collinear convolution integrals $J_{c}^{(J)}$ read
\bea
J_{c}^{(J)}[P_{2}V_{3}]=
\frac18 \int_{0}^{1}dy~\frac{\bar \phi_{2P}(y)}{y\bar{y}}\int
_{0}^{1}\frac{dx}{x\bar{x}}~\left\{  C_{+}^{(J)}(x,y)\int_{x}^{1}
du\frac{\Delta\Omega(u)}{u}+C_{-}^{(J)}(x,y)\int_{0}^{x}du\frac{\Delta \Omega(u)}{\bar{u}
}\right\}
\nonumber \\
+
\frac18\int_{0}^{1}dy~\frac{\Delta \phi_{2P}(y)}{y\bar{y}}\int
_{0}^{1}\frac{dx}{x\bar{x}}~\left\{  C_{+}^{(J)}(x,y)\int_{x}^{1}
du\frac{\bar\Omega(u)}{u}+C_{-}^{(J)}(x,y)\int_{0}^{x}du\frac{\bar \Omega(u)}{\bar{u}
}\right\}.
\label{def:JcP2V3}
\eea
Here we used convenient notation for the symmetric and antisymmetric combinations
\bea
\bar{\phi}(x)=\frac12(\phi(x)+\phi(\bar{x})), \quad \Delta{\phi}(x)=\frac12(\phi(x)-\phi(\bar{x})), \,  \quad  \bar{x}\equiv 1-x.
\label{def:symasym}
\eea
The hard kernels in Eq.(\ref{def:JcP2V3}) read 
\begin{equation}
C_{-}^{(1)}(x,y)=\frac{2\bar{x}}{\left(  xy+\bar{x}\bar{y}\right)  }%
-\frac{\bar{x}(x+\bar{y})}{\left(  xy+\bar{x}\bar{y}\right)  ^{2}}%
,~~~C_{+}^{(1)}(x,y)=\frac{2\bar{x}-1}{xy+\bar{x}\bar{y}}+\frac{\bar{x}%
(y-\bar{x})}{\left(  xy+\bar{x}\bar{y}\right)  ^{2}}.
\end{equation}%
\begin{equation}
C_{-}^{(2)}(x,y)=\frac{\bar{x}(x+\bar{y})}{\left(  xy+\bar{x}\bar{y}\right)
^{2}},~\ ~C_{+}^{(2)}(x,y)=-\frac{x(\bar{x}+y)}{\left(  xy+\bar{x}\bar
{y}\right)  ^{2}}.
\end{equation}%
The function $\Omega(u)$ is defined in Eq.(\ref{def:Omega}) in Appendix A . 

 For the convolution integral describing  $P_{3}V_{2}$  projection can be written as 
\bea
J_{c}^{(J)}[P_{3}V_{2}]= \frac{(-1)}{48}\int_{0}^{1}dx
\frac{\bar{\phi}_{2V}^{\bot}(x)}{x\bar{x}}
\int_{0}^{1}dy~
\frac{\tilde{C}^{(J)}_{\sigma}(x,y)\Delta\phi^{\sigma\prime}_{3P}(y)+C^{(J)}_{p}(x,y)\Delta\phi^{p}_{3P}(y)+{C}^{(J)}_{\sigma}(x,y) \Delta\phi^{\sigma}_{3P}(y)}{y\bar
{y}(xy+\bar{x}\bar{y})^{2}}
\nonumber \\
+\frac{(-1)}{48}\int_{0}^{1}dx\frac{\Delta{\phi}_{2V}^{\bot}(x)}%
{x\bar{x}}\int_{0}^{1}dy~\frac{\tilde{C}^{(J)}_{\sigma}(x,y)\bar{\phi}^{\sigma\prime}_{3P}(y)
+C^{(J)}_{p}(x,y)\bar{\phi}^{p}_{3P}(y)+{C}^{(J)}_{\sigma}(x,y) \bar{\phi}^{\sigma}_{3P}(y)}{y\bar
{y}(xy+\bar{x}\bar{y})^{2}},
\label{def:JcP3V2}
\eea
 where 
\begin{align}
   \tilde{C}^{(J=1)}_{\sigma}& =(y-\bar{y})(y-\bar{x}),  &\tilde{C}^{(J=2)}_{\sigma} &=(1+y\bar{y}-x\bar {x}-(x-y)^2), 
 \\
 C^{(J=1)}_{p}&=6(\bar{x}-y+2(y-\bar{y})(xy+\bar{x}\bar{y})),\,  &C^{(J=2)}_{p}&=6 (y-\bar{x}),
 \\ 
   {C}^{(J=1)}_{\sigma} &=4(\bar{x}-y),  \,   & {C}^{(J=2)}_{\sigma} & =4(y-\bar{x}).
\end{align}
The explicit expressions for  DAs  $\phi^{\sigma, p}_{3P}$ and $\phi_{2V}^{\bot}$ are given  in Eqs.(\ref{def:phis3P}),(\ref{def:phip3P}) and (\ref{mod:phiIIV}), 
the prime denotes  derivative with respect to  collinear fraction: $\phi'(x)\equiv d/dx \phi(x)$.
From Eq.(\ref{res:A0Jbot})  one can easily conclude that  the transverse amplitudes behave as 
\bea
\mathcal{A}_{Jc}^{\bot(0)}\sim  v^4\left(\frac{\Lambda}{m_c}\right)^3,
\label{scaleA0}
\eea
and  these contributions are suppressed compared to $\mathcal{A}_{1}^{\Vert}$.  On the other side,   amplitudes $\mathcal{A}_{J}^{\bot}[P_{3}V_{2}]$  include the so-called chiral enhanced  coefficient $\mu_P$, see Eq.(\ref{def:muK}), which is numerically large. Taking into account  the real value of the $c$-quark mass one finds that  $\mu_P/m_c\sim 1$ and  therefore  such corrections can provide a large effect.   

The convolution integrals $J_c^{(J)}$ have logarithmic  IR-divergencies associated with the endpoint regions $y\to 0,\, x\to 1$ and $y\to 1,\, x\to 0$.  These are the so-called  endpoint divergencies which indicate about the logarithmic overlap with the ultrasoft domain.  In order to single out  these  divergencies one needs to perform an expansion of the integrands in the corresponding  regions.  

Consider, for instance,  the  first integral  in Eq.(\ref{def:JcP3V2}) which has  the following integrand
\bea
F(x,y)=\frac{\bar{\phi}_{2V}^{\bot}(x)}{x\bar{x}} \frac{\tilde{C}^{(J)}_{\sigma}(x,y)\Delta\phi^{\sigma\prime}_{3P}(y)+\dots}{y\bar
{y}(xy+\bar{x}\bar{y})^{2}},
\label{def:F}
\eea
here the dots denote the other  terms in the numerator of  Eq.(\ref{def:JcP3V2}).
 Using  models of DAs  from Appendix~A one  easily finds  the following useful relations
\bea
&&  \lim_{x\rightarrow1}\bar{\phi}_{2V}^{\bot}(x)=-\bar{x} \bar{\phi}_{2V}^{\bot \prime}(1), \quad  
\lim_{x\rightarrow0}\bar{\phi}_{2V}^{\bot}(x)={x}\bar{\phi}_{2V}^{\bot\prime}(0),  
\\[2mm]
&&\lim_{y\rightarrow0}\Delta\phi^{p}_{3P}(y) =\rho_{-}^{K}~\frac32 \left( \alpha+\beta\ln y \right)\equiv \Delta\phi^{p}_{3P}(y\sim 0),
\quad
 \alpha=1+21b_2,\, \beta=1+6b_2, 
 \label{log0}
\\[2mm]
&& \lim_{y\rightarrow0}  \Delta\phi^{\sigma\prime}_{3P}(y) =  6\Delta\phi^{p}_{3P}(y\sim 0),\quad
\lim_{y\rightarrow1} \Delta \phi^{\sigma\prime}_{3P}(y) = -6\Delta\phi^{p}_{3P}(y\sim 1),
\\[2mm]
&&\lim_{y\rightarrow1}\Delta\phi^{p}_{3P}(y) =-\rho_{-}^{K}~\frac32\left( \alpha+ \beta\ln \bar{y}\right)\equiv \Delta\phi^{p}_{3P}(y\sim 1).
\label{log1}
\eea
Hence the expansion of integrand in Eq.(\ref{def:F}) in the endpoint regions gives
\bea
&&
\left. F(x,y)
   \right\vert_{y\sim \bar{x}\rightarrow 0}
=(-1)^{J+1}4 \bar{\phi}_{2V}^{\bot\prime}(1)\,
\frac{ 6 \Delta\phi^{p}_{3P} (y\sim0)}{(y+\bar{x})^{2}}\equiv F(x\sim 1, y\sim 0) ,
 \label{div1}
 \\[2mm]
&& \left. F(x,y)
   \right\vert_{x\sim \bar{y}\rightarrow 0}
=(-1)^{J+1}4 \bar{\phi}_{2V}^{\bot\prime}(0)\, 
\frac{6\Delta\phi^{p}_{3P}(y\sim1)}{(x+\bar{y})^{2}}\equiv F(x \sim 0, y\sim 1) .
 \label{div2}
 \eea
The corresponding convolution  integral in Eq.(\ref{def:JcP3V2})  can be  rewritten as a sum of the regular and singular  terms
\bea
\int_0^1 dx  \int_0^1 dy  F(x,y) = I_{\text{reg}}+I_{\text{sing }} ,
\eea
with 
\bea 
I_{\text{reg}}= \int_0^1 dx  \int_0^1 dy \{ F(x,y)-F(x\sim 1, y\sim 0)-F(x \sim 0, y\sim 1) \}, 
\nonumber\\ 
I_{\text{sing}}= \int_0^1 dx  \int_0^1 dy \{  F(x\sim 1, y\sim 0)+F(x \sim 0, y\sim 1) \},
\label{Ising}
\eea
where only the  integrals $I_{\text{sing}}$ are IR-divergent.  In order to regularise  them  we apply  analytic regularisation  modifying  the heavy quark propagator.  We imply that the regularisation is introduced { \it after}  differentiation with respect  to
relative momentum $\Delta_\top$ ( as required by projection on  quarkonium  $P$-wave state)
\bea
\frac{1}{[m_c^2(x y+ \bar x \bar y)]^n}\to \frac{\nu^{2\varepsilon}}{[m_c^2(x y+ \bar x \bar y)]^{n+\varepsilon}},
\eea 
where $\nu$ is the renormalisation scale. With such regulator one obtains  
\begin{equation}
I_{\text{sing}}=(-1)^{J+1} 4 \bar{\phi}_{2V}^{\bot\prime}(1)~t^{\varepsilon}\int_{0}^{1}dx\int_{0}^{1}dy~
\left\{ \frac{ 6 \Delta\phi^{p}_{3P}  (y\sim0)}{(y+\bar{x})^{2+\varepsilon}}-
\frac{ 6 \Delta\phi^{p}_{3P}  (y\sim1)  }{(\bar{y}+x)^{2+\varepsilon}}\right\},
\end{equation}%
where $t\equiv \nu^2/m_c^2$ and we used that $ \bar{\phi}_{2V}^{\bot\prime}(1)=-\bar{\phi}_{2V}^{\bot\prime}(0)$.  
 A simple but lengthy  calculation yields 
\begin{align}
I_{\text{sing}}=& (-1)^{J+1} \bar{\phi}_{2V}^{\bot\prime
}(1)72\rho_{-}^{K}
\nonumber \\
&\times\left(  -\frac{\beta}{\varepsilon^{2}}-\beta\frac{\ln
t}{\varepsilon}-(\alpha-\beta)\frac{1}{\varepsilon}-\beta\frac{1}{2}\ln^{2}t-(\alpha-\beta)\ln t+\beta\frac{\pi^{2}}{12}-(\alpha-\beta)-\alpha\ln2 \right).
\label{IRlogs}
\end{align}%
The double IR-pole in $1/\varepsilon$ arises due to the presence of  logarithms $\ln y$ and $\ln \bar{y}$  in Eqs.(\ref{log0}) and (\ref{log1}).   
Hence for the first integral from Eq.(\ref{def:JcP3V2}) we obtain
\begin{align}
J_{c1}^{(J)}[P_{3}V_{2}] =&-\frac1{48}\int_0^1dx \int_0^1dy F(x,y)
\nonumber \\
=&(-1)^{J}  \bar{\phi}_{2V}^{\bot\prime}(1)\frac32 \rho_{-}^{K} \left(   -\frac{\beta}{\varepsilon^{2}}-\beta\frac{\ln
t}{\varepsilon}-(\alpha-\beta)\frac{1}{\varepsilon}-\beta\frac{1}{2}\ln^{2}t-(\alpha-\beta)\ln t\right) + \dots,
\label{res:Jc1div}
\end{align}
where dots denote the remnant finite terms. The same technique  can also be used   for other convolution integrals in Eqs.(\ref{def:JcP2V3}) and (\ref{def:JcP3V2}). These integrals also have  IR-divergencies  which produce  double and single poles in $1/\varepsilon$.  	

 A study of  structure of the divergent integrals  can be helpful in order  to identify an operator  which can be associated with the IR-divergencies.  The intermediate gluons  in the diagrams in Fig.\ref{diagrams} have momenta $yk+\bar{x}p$ and  $\bar{y}k+xp$.  Hence in the regions  $y\sim \bar{x}\rightarrow 0$ or $x\sim \bar{y}\rightarrow 0$  one of the gluons  has a very small momentum  while the second gluon still  has the hard momentum.  It is natural to assume that  gluon with  the small momentum is ultrasoft, i.e. in the endpoint regions  we have $yk+\bar{x}p\sim m_cv^2$ or $\bar{y}k+xp\sim m_cv^2$ which is equivalent  to $y\sim \bar{x}\sim v^2$ or $x\sim \bar{y}\sim v^2$.  The interactions of such  ultrasoft  gluons with a soft heavy quark  $p_Q\sim m_cv$  does not change its virtuality.  Therefore in the endpoint domain the  momentum of the virtual  heavy quark in diagrams  in Fig.\ref{diagrams}  is soft.  The corresponding  propagators  yield  combinations $(y+\bar{x})^{-2}$  or  $(x+\bar{y})^{-2}$ in the kernels of Eqs.(\ref{div1}) and (\ref{div2}) and these terms  produce  the  IR-divergencies  in the  convolution integrals.  
 
 Therefore in the endpoint regions  the hard subprocess  is different and can be described by  the hard annihilation of  heavy quark-antiquark pair into  the  light quark-antiquark pair  with light-like momenta: $c\bar{c}\to g^*\to q+\bar{q}$   or  $c\bar{c}\to g^*\to s+\bar{s}$. Since the annihilation produces  only one hard gluon the corresponding heavy quark-antiquark pair must be  in the colour-octet state.  This allows one to  conclude that a colour-octet matrix element  must be added into the consideration  in  order to  explain  IR-divergencies of the colour-singlet contribution.  It is obvious  that such octet contribution must have the same  behaviour in $v$ as the singlet one in Eq.(\ref{scaleA0}).  
  
  The  mixing of singlet and octet mechanisms in exclusive decays  within the effective theory framework has already been studied  in Ref.\cite{Beneke:2008pi}.  
  In present case  the situation is similar but a bit more   complicated  from the technical point of view  because of  double IR-poles, see Eq.(\ref{IRlogs}).   
  In  the realistic  world  the colour-octet  contribution is non-perturbative because of relatively small charm mass.  However in the next section  we consider corresponding matrix element in the Coulomb limit  which allows one to  perform  calculations within the  pNRQCD framework. Such consideration allows  one  explicitly to verify  the  correspondence  of  divergencies between  the colour-singlet and colour-octet  terms.   If  IR-poles in colour-singlet matrix element  are reproduced as UV-poles of the colour-octet  contribution then  the IR-poles can be absorbed  into the renormalisation of the colour-octet matrix element.  
  
Let us  consider  some  qualitative arguments  based on the spin symmetry of the effective field theory in the  limit $m_c\to \infty$.   It is  well known that  the  HQSS  provides  approximate relations between matrix elements for the various states of a given radial and orbital excitation of heavy quarkonium. 
 The violation of  the heavy-quark spin symmetry  related with the higher order  terms in effective Lagrangian  suppressed by powers of $v$.  The example of such relations for the wave functions are well known \cite{Bodwin:1994jh}  and used in  Eqs.(\ref{def:O3P1}-\ref{<O3PJ>}).  Despite the colour-octet operators are more complicated  the effective heavy quark Lagrangian is the same  and  this also can provide an approximate relations between the various octet matrix elements.           
  
The  relevant for our case  hard subprocess is $c\bar{c}\to g^*\to q+\bar{q}$  and  hard  factorisation   yields  the  four-quark operators  
  \bea
 C_h\,  \bar {q}\gamma_\sigma t^a q\, \,  \chi_\omega^{\dag} \gamma^\sigma_\top t^a\psi_\omega , 
  \eea  
  where $C_h$ is  the hard coefficient function, $t^a$ denotes  the SU(3) color matrices, $q$ denotes the light quark field,   $\psi _{\omega }$ and $\chi _{\omega }^{\dag }$ denote  quark and antiquark 
 four-component spinors in the NRQCD, see more details in Appendix~A.  Then  the  colour-octet amplitude is schematically given by the matrix element 
  \bea
  \mathcal{A}^{\bot(8)}_J =C_h\,  \langle K^*\bar K\vert \bar {q}\gamma_\sigma t^a q\, \,  \chi_\omega^{\dag} \gamma^\sigma_\top t^a\psi_\omega\vert \chi_J(\omega)\rangle.
  \label{col-oct}
  \eea  
 According to NRQCD counting rules,  the  bilinear heavy quark operator  is of order  $v^3$.   In order to get a contribution of order $v^4$ which can mix with the colour-singlet  contribution in (\ref{scaleA0})  one  needs  an interaction   of order $v $.  In pNRQCD  Lagrangian  such interaction  is  only described  by   chromoelectric dipole vertex $\sim \psi^\dag_\omega (x) \vec{x}\cdot\vec{E}(t) \psi_\omega(x)$  which is not sensitive to the heavy quark spin.  Therefore we can conclude  that  HQSS can also relate  the matrix elements (\ref{col-oct}) with the different $J=1,2$.  In Sec.~4 it  will be shown that in the weak coupling limit $|p_{us}|\gg \Lambda$ this yields  
 \bea
 \mathcal{A}^{\bot(8)}_{J=1} =-\frac{1}{\sqrt{2}}\mathcal{A}^{\bot(8)}_{J=2}.
 \label{SSrel}
 \eea       
  up to higher order corrections in small velocity $v$.   The next important  step is the assumption that  at  given order   the total result for the physical amplitude  is only given by the sum of the singlet and octet amplitudes.  Then the various factorisation scales  which appears in these contributions  must cancel  in the sum  
  \bea
\mathcal{A}^{\bot}_{J}= \mathcal{A}^{\bot(0)}_{J}+\mathcal{A}^{\bot(8)}_{J} .
\label{Aperp: fact}
 \eea      
 Such compensation in some sense is equivalent to a cancellation  of  singularities  in the {\it rhs} of  Eq. (\ref{Aperp: fact}),  therefore  this   implies  that  divergent integrals  in the colour-singlet amplitudes must also satisfy to relation  (\ref{SSrel}).  Then  the  hard contributions which violate spin-symmetry   relations must be  well defined, i.e. they are free from IR-singularities  and therefore can be computed unambiguously.  A similar  situation  takes plays  in B-decays \cite{Beneke:2000wa, Beneke:2001ev}.  
 We can relate the amplitudes with the different $J$ using  the so-called  physical subtraction scheme \cite{Beneke:2000wa, Beneke:2001ev}.  Using Eqs.(\ref{SSrel}) and (\ref{Aperp: fact}) in order to exclude  colour-octet  amplitude $\mathcal{A}^{\bot(8)}_{J}$ one obtains
\bea
\mathcal{A}^{\bot}_{1}+ \frac{1}{\sqrt{2}} \mathcal{A}^{\bot}_{2}= \mathcal{A}^{\bot(0)}_{1}+\frac{1}{\sqrt{2}}\mathcal{A}^{\bot(0)}_{2} .
\label{A12:rel}
\eea   
The  combination of the colour-singlet  amplitudes in the {\it rhs} of this equation must be  well defined  since the {\it lhs} is free from any divergencies.  This point can be easily verified  using  results in Eqs.(\ref{res:A0Jbot}).  Performing the required analytical calculations  we indeed obtain  that  the combination $\mathcal{A}^{\bot(0)}_{1}+\frac{1}{\sqrt{2}}\mathcal{A}^{\bot(0)}_{2} $ is free from the endpoint divergencies. This observation  supports the factorisation formula  suggested in Eq.(\ref{Aperp: fact}).  Notice that this compensation works  independently for  two different collinear operators describing $P_2 V_3$ and $P_3V_2$ projections.    Relation (\ref{A12:rel})  is one of the main  results of this work and it  will  be used in our  phenomenological analysis  in Sec.\ref{phen}.  

At the end of this section let us  provide the  analytical results  for the collinear integrals which define the  symmetry breaking contributions in Eq.(\ref{A12:rel})         
\begin{align}
 \mathcal{A}_{1c}^{\bot(0)}+\frac{1}{\sqrt{2}}\mathcal{A}_{2c}^{\bot(0)}  & \equiv  \Delta\mathcal{A}_{c}^{\bot(0)}
\nonumber \\  &
  =\frac{~i\left\langle \mathcal{O}(^{3}P_{J})\right\rangle }{m_c^{3}}
\frac{~f_{P}f^{\Vert}_{V}m_{V}}{m_c^{3}}\left(  \frac{\pi\alpha_{s}}{N_{c}}\right)
^{2}~C_{F}~\sqrt{2}\left(  J_{c}^{(1)}[P_{2}V_{3}]+J_{c}^{(2)}[P_{2}%
V_{3}]\right)  
\nonumber \\ & 
 +\frac{i\left\langle \mathcal{O}(^{3}P_{J})\right\rangle }{m_c^{3}}%
\frac{f_{P}\mu_{P}f_{V}^{\bot}~}{m_c^{3}}\left(  \frac{\pi\alpha_{s}}{N_{c}%
}\right)  ^{2}C_{F}\sqrt{2}\left(  J_{c}^{(1)}[P_{3}V_{2}%
]+J_{c}^{(2)}[P_{3}V_{2}]\right).
\label{def:DA0}
\end{align}
Using the models of DAs from  Appendix~A  we obtain
\begin{align}
&J_{c}^{(1)}[P_{2}V_{3}]+J_{c}^{(2)}[P_{2}V_{3}]= \frac 32 b_{1}\left\{  \frac{9}%
{4}\left(  8-\pi^{2}\right)  +a_{2V}^{\Vert}\frac{9}{16}\left(  11\pi
^{2}-108\right)  \right\}
\nonumber \\
& + \frac 32~b_{1}\lambda_{s}^{+}~\left[  \frac{9}{2}\left(  8-3\zeta(3)+\pi
^{2}(1-2\ln2)\right)  +9a_{2V}^{\bot}\left(  9(1-\zeta(3))+\pi^{2}\left(
\frac{17}{4}-6\ln2\right)  \right)  \right] 
 \nonumber \\
& + \frac 32 b_{1}\lambda_{s}^{-}a_{1V}^{\bot}\frac{27}{4}(6\zeta(3)-24+\pi^{2}%
(4\ln2-1))
+ \frac 32 a_{1V}^{\Vert}\left\{  \frac{3}{4}(8-\pi^{2})+b_{2}\frac{9}{8}\left(
11\pi^{2}-108\right)  \right\}
\nonumber \\
& + \frac 32~\lambda_{s}^{-}\left\{  -\frac{3}{4}\left[  6\zeta(3)+4\pi^{2}%
(\ln2-1)\right]  -\frac{9}{4}b_{2}\left(  12\zeta(3)-40+\pi^{2}\left(
8\ln2-3\right)  \right)  \right. 
\nonumber \\
& \left.  -a_{2V}^{\bot}\frac{3}{2}\left[  18\zeta(3)-40+\pi^{2}%
(12\ln2-7)+\frac{3}{2}b_{2}\left(  72\zeta(3)+300+\pi^{2}(48\ln2-73)\right)
\right]  \right\}
 \nonumber \\
& + \frac 32\left(  -\frac{9}{8}\right)  \lambda_{s}^{+}a_{1V}^{\bot}\left\{
~2\left(  6-\zeta(3)-8+\pi^{2}(4\ln2-3)\right)  \right. 
\left. +b_{2}\left[  84+\pi^{2}(48\ln2-51)+72\zeta(3)\right]  ~\right\}  ,
\label{res:sumJcP2V3}
\end{align}
\begin{align}
J_{c}^{(1)}[P_{3}V_{2}]+J_{c}^{(2)}[P_{3}V_{2}]  & =-\frac98 \rho_{-}^{K}\left\{
2\pi^{2}+a_{2V}^{\bot}3\left(  20-\pi^{2}\right) 
 +b_{2}\left(  20-\pi^{2}+a_{2V}^{\bot}\left(  39\pi^{2}-360\right)
\right)  \right\}  ,
\label{res:sumJcP3V2}
\end{align}
where we assume that all  parameters of DAs depend on the factorisation scale $\mu$.  From these results one can also see that in the limit of exact $SU(3)$ symmetry  expressions (\ref{res:sumJcP2V3}) and (\ref{res:sumJcP3V2}) vanish as it must be. 

\subsection{Soft-overlap colour-singlet  contribution to amplitudes $\mathcal{A}_{1,2}^{\bot}$ }

There is one more contribution which can provide a significant effect and therefore must be taken into account. This contribution  appears due to long distance  interactions between the outgoing partons and can be  associated with the typical hadronic scale $\Lambda$.  In this case  heavy quark and antiquark  annihilate at short distances into the light quark-antiquark  pair with the hard-collinear momenta $p_{hc}^2\sim m_c\Lambda$.  The light-cone fractions  of these  momenta are large and close to the total momenta of  outgoing mesons.  In order to produce final hadronic states  the hard-collinear particles interact with the soft and collinear  particles.  Corresponding subprocess depends on the hard-collinear and soft  virtualities which are of order $m_c\Lambda$ and $\Lambda^2$, respectively.  Such contribution can  be described  as  a  matrix element within the soft collinear effective theory (SCET) framework. 

\begin{figure}[ptb]%
\centering
\includegraphics[width=4.7335in]%
{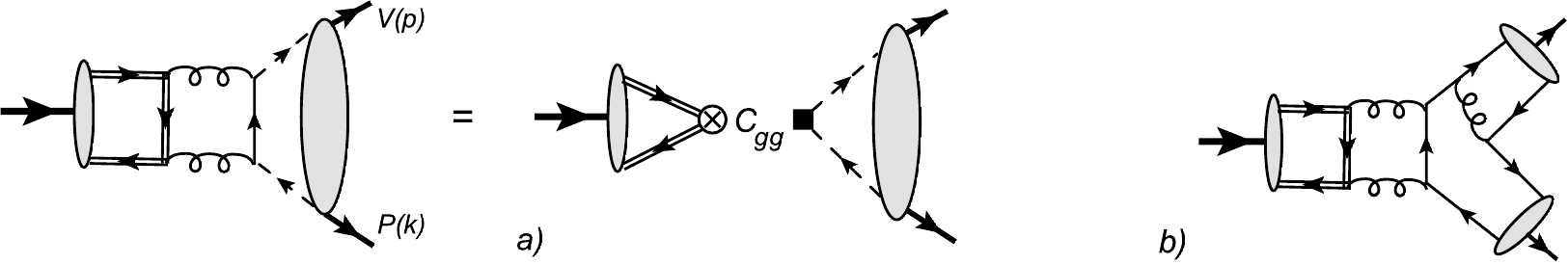}%
\caption{  $a)$ An example of the  one-loop diagram and schematic factorisation for the  soft-overlap colour-singlet contribution. The black square denotes the operator vertex and  dashed fermion lines with the blob  denote  SCET matrix element defined in Eq.(\ref{def:scetme}).  The crossed vertex with the blob describe the NRQCD matrix element.  $b)$ The higher order  perturbative diagram which must have  IR-singularities associated with the  soft-overlap configuration. }
\label{fig_soft-ovp-sing}
\end{figure}

Corresponding diagrams  are  schematically shown in Fig.\ref{fig_soft-ovp-sing}$(a)$, the  dashed lines denote the hard-collinear particles which are attached to the blob denoting  the SCET matrix element.   We assume that  the  hard-collinear scale $m_c\Lambda$ is not large  and  consider  the SCET matrix elements as non-perturbative objects.  If in the limit $m_c\to\infty$  these matrix elements  are  of order $(\Lambda/m_c)^3$  then  the  soft-overlap amplitude  is of the same order  as  the hard one, see Eq.(\ref{scaleA0}).  This can be directly verified  in SCET-II \cite{Bauer2001} by construction of the  relevant $T$-products or  by direct computation  of the higher-order diagrams as  in Fig.\ref{fig_soft-ovp-sing}$(b)$. Such diagrams  must have specific collinear endpoint singularities  which can be  associated with the SCET matrix elements.  Such  calculations are known  for  quite similar space-like  amplitude  describing  the process $\gamma^*\rho\to \pi$, see e.g. Refs.\cite{Chernyak:1983ej,Manohar:2006nz }. The detailed analysis of this point  is  quite complicated and  we  accept  that  power  behaviour  of  the soft-overlap contribution like  $(\Lambda/m_c)^3$   as a reliable assumption.   
 
  The  soft-overlap matrix elements describe a configuration when the  outgoing hard-collinear partons carry almost  total hadronic momentum.   Such situation  can be interpreted as a soft-overlap of  the final hadronic states.  For  space-like  form factors such scattering configuration  is also known  as a Feynman mechanism \cite{Feynman:1973xc} and corresponding effect has been  studied  long time ago  with the help of the light-front wave functions  \cite{Isgur:1984jm}.  
   
 The hard coefficient functions $C_{gg}$  are given by the sum of the one-loop diagrams like one  in Fig.\ref{fig_soft-ovp-sing}$(a)$.  The resulting expression for the  colour-singlet soft-overlap amplitudes can be written as 
\begin{align}
 \left. \mathcal{M}_{\chi_{cJ}\to \bar{K}K^*}\right|_{\text{Fig.}\ref{fig_soft-ovp-sing}a}\simeq &\frac{~i\left\langle \mathcal{O}(^{3}%
P_{J})\right\rangle }{m_c^{3}}~C_{gg}^{(J)}~\{  \delta_{J1}\,  i\epsilon^{\perp}_{\rho\sigma}  \epsilon_{\chi}^{\sigma}+ \delta_{J2}\, (\epsilon_{\chi})_{\rho\sigma} \bar{n}^\sigma \}
\left\langle \bar{K}K^{\ast}\right\vert \mathcal{O}_{SCET} \left\vert 0\right\rangle,
\label{def:Msov}
\end{align}
  where $\delta_{Ji}$ is Kronecker symbol,  the  matrix element of the  two-particle  SCET operator  $\mathcal{O}_{SCET}$ is  defined as 
 \begin{equation}
\left\langle \bar K(k) K^*(p, e^*)\right\vert \bar{s}_{\bar{n}}(0)W_{\bar{n}}\gamma_{\bot}^{\alpha}W^\dag_n s_{n}(0)-\bar{q}_{n}(0)W_{n}\gamma_{\bot}^{\alpha}W^\dag_{\bar{n}} q_{\bar{n}}(0)\left\vert 0\right\rangle
=i\varepsilon^{\bot}_{\alpha \beta}(e_{V}^{\ast})^\beta m\left(  f_{PV}^{s}-f_{PV}^{q}\right).
\label{def:scetme}
\end{equation}  
Here the quark fields $\psi_n$,  $\psi_{\bar{n}}$  ($\psi=s,q$) and $W_{n,\bar{n}}$ denote the hard-collinear SCET fields and  corresponding  hard-collinear Wilson lines
\bea
\Dsl{n} \psi_n(x)=0, ~  W_n=P\exp\left\{ ig \int_{-\infty}^0 ds ~\bar{n}\cdot A(s\bar{n})  \right\},
\label{def:Wn}
\eea
and similarly for the light-cone sector associated with  $\bar{n}$.  

The  form factors $f_{PV}^{s,q}$  describe transition of the  hard-collinear quark-antiquark pair to the  final hadronic state  within the SCET framework.   The relative sign  minus in Eq.(\ref{def:scetme}) can be understood as an consequence of  $C$-parity of the initial state. We see  that in the exact $SU(3)$  limit such matrix element vanishes as it must be.  The definition (\ref{def:scetme}) is process independent,  the similar  matrix element may also appear in other hard reactions, for instance,   in the wide angle scattering   $\gamma\gamma\to \bar{K}K^{\ast}$ at large energy and momentum transfer.   The  different combination of these form factors  can also appear in the process  $e^+e^-\to\gamma^*\to \bar{K}K^{\ast}$.  
Using  Eqs.(\ref{def:Msov}) and (\ref{def:scetme}) one can easily find  corresponding contributions to the colour-singlet amplitudes $\mathcal{A}_{J}^{\perp(0)}$\begin{equation}
\mathcal{A}_{Js}^{\perp(0)}=\frac{~i\left\langle \mathcal{O}(^{3}P_{1}%
)\right\rangle }{m_c^{3}}~C_{gg}^{(J)}~\left(  f_{PV}^{s}-f_{PV}%
^{q}\right)  .
\label{def:A0Js}
\end{equation}

The computation of the hard coefficients $C_{gg}^{(J)}$ is strightforward: one has to compute the box diagrams as in Fig.\ref{fig_soft-ovp-sing}$(a)$ in the appropriate kinematics.  The required  one-loop integrals are similar to the integrals studied for $\chi_{J}\to e^+e^-$  decays. We borrow the results from  Refs.\cite{Kivel:2015iea}   adding the colour factor and QCD couplings.  These integrals have IR-divergencies which are regularised by dimensional regularisation $D=4-2\varepsilon$.  Using  $\overline{MS}$-scheme one obtains
\begin{equation}
C_{gg}^{(J=1)}=\alpha_{s}^{2}\frac{C_{F}}{N_{c}}\frac{1}{\sqrt{2}}%
\left( -\frac{1}{\varepsilon}-\ln\frac{\mu^{2}}{m_c^{2}}-2\ln2\right),\ C_{gg}^{(J=2)}=~\alpha_{s}^{2}%
\frac{C_{F}}{N_{c}}\left\{\frac{1}{\varepsilon}+ \ln\frac{\mu^{2}}{m_c^{2}}+\frac{2}{3}\left(  \ln
2-1+i\pi\right)  \right\} ,
\label{res:Cgg}
\end{equation}
where $\mu$  is the factorisation scale.   The IR-singularities in the hard loop  corresponds  to the integration  domain where one of the gluons becomes ultrasoft.   Corresponding IR-poles can be again absorbed into the colour-octet matrix element which  will be discussed  this in Sec.~4.  The total result  for a soft-overlap amplitude  is  also given by the sum of  colour-singlet and colour-octet  matrix elements.  

The soft-overlap colour-octet contribution also satisfies  Eq.(\ref{SSrel}) as a part of the total colour-octet amplitude. Therefore  one can also apply the same arguments and obtain the relation like  Eq.(\ref{A12:rel}) which allows one to define the corresponding HQSS breaking terms.  Using Eq.(\ref{res:Cgg})  we  obtain
\begin{equation}
\mathcal{A}_{1s}^{\perp(0)}+\frac{1}{\sqrt{2}}\mathcal{A}_{2s}^{\perp (0)}\equiv\Delta \mathcal{A}_{s}^{\bot(0)} =\frac
{i\left\langle \mathcal{O}(^{3}P_{2})\right\rangle }{m_c^{3}}\left( f_{PV}^{s}-f_{PV}^{q}\right)  \alpha_{s}^{2}\frac{C_{F}}{N_{c}}\frac
{\sqrt{2}}{3}\left(  -1-2\ln2+i\pi\right)  .
\label{res:A0s}
\end{equation}
We again confirm  that IR-poles and factorisation scale  cancel as it is expected. 
Notice that this contribution has imaginary part which is related to the two-gluon intermediate cut in the loop diagram. This contribution is also of  order  $\alpha^2_s$   and  have the same power counting behaviour as $\mathcal{A}_{\bot}^{(0)}$  in Eq.(\ref{def:DA0}).  If the  value of the SCET matrix elements  $\sim ( f_{PV}^{s}-f_{PV}^{q})$ is sufficiently  large  then this  contribution cannot be neglected. 

\section{Colour-octet contributions  in the Coulomb limit}

For a realistic charmonium   colour-octet matrix elements can be computed only within a non-perturbative framework. However in order to study certain properties of the NRQCD  matrix elements it could  be useful to consider a special limit, also  known as the  Coulomb limit, when  the ultrasoft scale is sufficiently large.   In such limit   the ultrasoft scale is a larger  than  the typical hadronic scale   $m_cv^2\gg \Lambda$ and quarkonium state can be considered as a weakly  bound state with  the binding energy $E\sim m_cv^2$.  Important  point is that the perturbation theory can be used  for  calculations associated with the ultrasoft scale.   The  standard  framework includes:  factorisation of hard modes and transition to NRQCD,  the integration over the soft and potential  gluons  and transition  to  pNRQCD  which only contains  potential heavy quarks and ultrasoft gluons as degrees of freedom.  Such  picture of course cannot  provide  reliable estimates for realistic charmonia but it  allows one to study a structure of the infrared divergencies which are related with the mixing of  colour-singlet and -octet operators.  Our aim is to show that the colour-octet matrix elements  satisfy to  Eq.(\ref{SSrel}) in the Coulomb limit and to study  UV- and IR-singularities  in  the  colour-singlet  and -octet contributions.   

\begin{figure}[ptb]%
\centering
\includegraphics[
width=4.0in
]%
{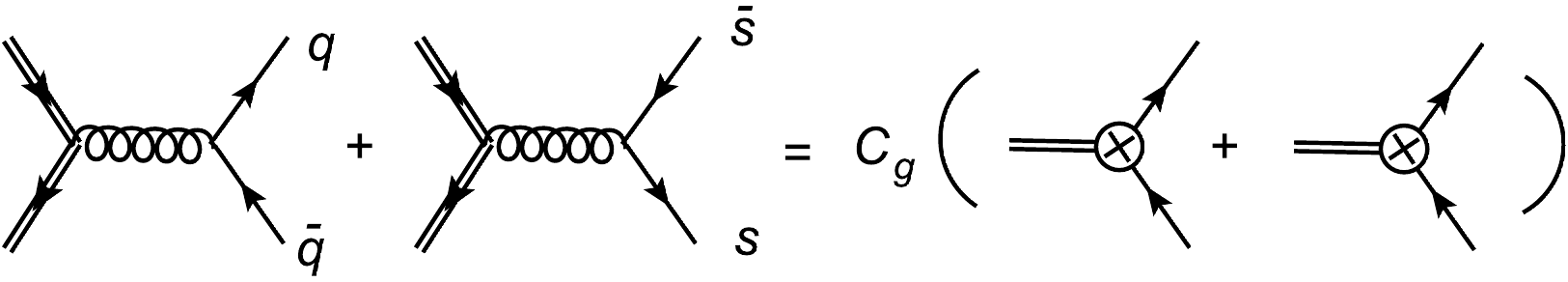}%
\caption{ The hard factorisation  associated with the  annihilation $c\bar{c}\to g^*\to q(s)+\bar{q}(\bar{s})$. The crossed circle  denotes the vertex of the four-fermion operator and the attached double line describes the heavy quark-antiquark pair. }%
\label{fig_h-fact}%
\end{figure}

In our  case the  factorisation of  hard modes  is described  by the tree level diagrams associated with the subprocess $c\bar{c}\to g^*\to q+\bar{q}$   or  $c\bar{c}\to g^*\to s+\bar{s}$  as shown graphically in Fig.~\ref{fig_h-fact}. Two  diagrams correspond to the two different regions $y\sim \bar{x}\sim v^2$ or $x\sim \bar{y}\sim v^2$ in the collinear integrals. 
Hence for the octet amplitudes we get 
\begin{equation}
i\mathcal{M}_{\chi_{cJ}\to K^*\bar{K}}^{(8)}=C_{g}\left\langle K^{\ast}\bar{K} \right\vert
\{\bar{s}_{\bar{n}}(0)W_{\bar{n}}\gamma_{\bot}^{\alpha}
t^{a}W^\dag_n s_{n}(0)+\bar{q}_{n}(0)W_n\gamma_{\bot}^{\alpha}t^{a}W^\dag_{\bar n}q_{\bar{n}}(0)\}
 \chi_{\omega}^{\dag}\gamma_{\top}^{\alpha}t^{a}\psi_{\omega}
\left\vert \chi_{cJ}\right\rangle ,
\label{A8hard}
\end{equation}
with the hard coefficient function 
\begin{equation}
C_{g}=\frac{i\alpha_{s}(\mu)\pi}{m_c^{2}}.
\end{equation}%
In Eq.(\ref{A8hard}) we also use  notation for the collinear fields and Wilsons lines  as in Eq.(\ref{def:Wn}).   

The next step is  transition  to  pNRQCD. 
To our accuracy the  matching of the NRQCD operator   in  (\ref{A8hard})   onto  pNRQCD operator  is trivial 
\bea
  \chi_{\omega}^{\dag}\gamma_{\top}^{\alpha}t^{a}\psi_{\omega}\vert_{NRQCD}=\chi_{\omega}^{\dag}\gamma_{\top}^{\alpha}t^{a}\psi_{\omega}\vert_{pNRQCD}\, .
  \eea
 Therefore we can easily pass to a calculation of the matrix element  (\ref{A8hard})  in pNRQCD.

\begin{figure}[ptb]%
\centering
\includegraphics[
width=4.0in
]%
{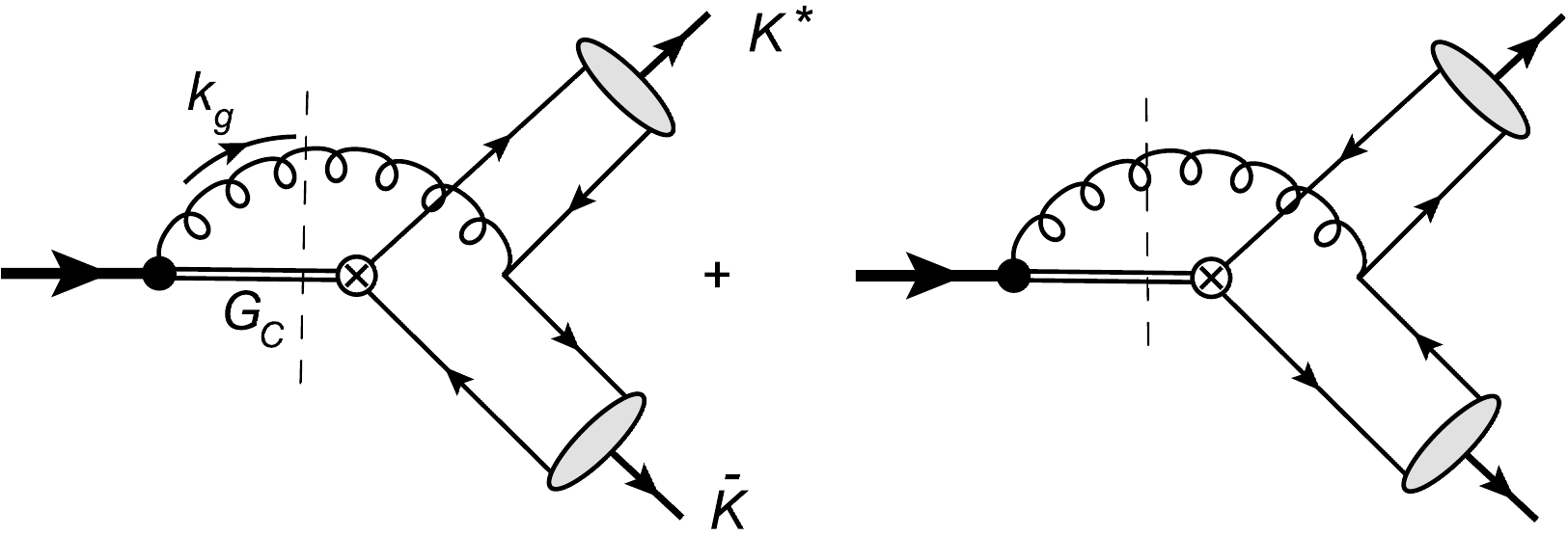}%
\caption{ The  diagrams in pNRQCD describing  the colour-octet matrix elements. The double line  denotes the colour-octet  Coulomb propagator $G_C$.  The dashed line shows the cut associated with  the imaginary part. }%
\label{fig_pNRQSD-diag}%
\end{figure}

\begin{figure}[ptb]%
\centering
\includegraphics[
width=4.0in
]%
{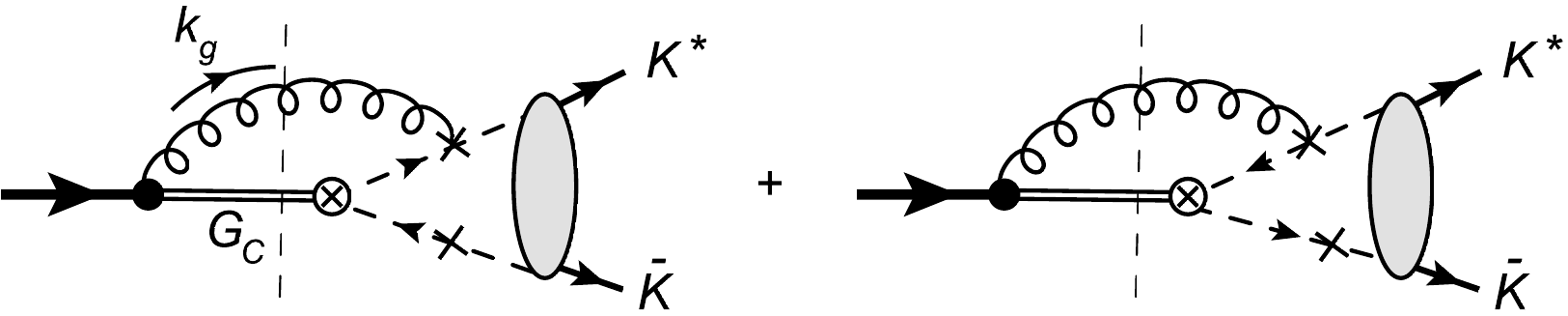}%
\caption{ The one-loop pNRQCD diagrams describing the  colour-octet matrix elements  with the soft-overlap matrix element. The crossed quark lines  indicate all possible attachments of the ultrasoft gluon.  }%
\label{fig_softoverlap}%
\end{figure}

To the leading-order  accuracy in $\alpha_s$  one can consider   two different sets of  Feynman diagrams: tree level and one-loop graphs which  
are shown in Fig.~\ref{fig_pNRQSD-diag} and Fig.~\ref{fig_softoverlap}.
   
   The tree  diagrams in Fig.~\ref{fig_pNRQSD-diag} have the following structure.  The  initial $P$-wave bound state  decays through the chromoelectric dipole interaction into ultrasoft gluon and bound  quark-antiquark pair in the colour-octet state. The  interaction vertex is suppressed by power of the small velocity $v$  therefore  the total octet contribution is of order  $v^4$ (remind,  the $S$-wave vector operator in (\ref{A8hard}) is of order  $v^3$).  The  colour-octet  quark-antiquark pair  propagates a distance $\sim 1/(m_cv^2)$ and annihilates into light  quark-antiquark pair with momenta of order $k$ and $p$. The colour-octet  propagator is described by the non-relativistic Coulomb Green function $G_C$.  The virtual ultasoft gluon  also creates  the light quark-antiquark pair  which together  with the collinear  quark-antiquark  provide  a  collinear operator describing  a long distance overlap with outgoing mesonic states. 
  Since $p_{us}^2\gg \Lambda^2$  the ultrasoft  particles  still  can be matched onto collinear degrees of freedom. However the corresponding collinear  fractions are small, of order $v^2$ as it follows from the momentum conservation.  Therefore corresponding collinear matrix elements describe an asymmetric collinear configurations where one  parton carries the small collinear  fraction $x\sim v^2$.  This is exactly the endpoint configuration which provides the IR-singularities in the colour-singlet matrix element.  The resulting  expression for such diagram must be expanded with respect to small collinear fractions keeping only those terms which  provide  contribution  of order  $v^4$.  The long distance dynamics associated with the hadronic scale $\Lambda$ is still described by the DAs  originating  from  the  collinear matrix elements.  Such situation is a  consequence  of presence of the two well separated scales $m_cv^2\gg \Lambda$  in  the Coulomb limit. 
   
 The one-loop diagrams  in Fig. \ref{fig_softoverlap} describe the colour-octet contribution associated with the soft-overlap amplitude.    
   In this case the ultrasoft gluon  interacts with the collinear quark or antiquark creating  the colourless quark-antiquark  operator.  
    Further interactions  of the hard-collinear particles  is only  associated with the typical hadronic scale  $\Lambda$  and  described as  matrix elements of the SCET operator which is  shown by  blob in 
   Fig.\ref{fig_softoverlap}.  As in Sec.~3 we consider  these matrix elements  as  non-perturbative quantities.    
   Obviously,  such contribution  also  is  of order $v^4$.  
     
   Technically  the calculation of the pNRQCD diagrams in Fig.\ref{fig_pNRQSD-diag}   is  similar  to  calculation  in Ref.\cite{Beneke:2008pi}  and useful technical details can be found in this work.  The analytical expression  for the colour-octet amplitude  can  be written as     
\begin{align}
\left.\mathcal{M}_{\chi_{cJ}\to K^*\bar{K}}^{(8)} \right |_{\text{Fig.}\ref{fig_pNRQSD-diag}}& =
-\frac{2 \pi^{2} }{m_c^{2}}\alpha_{s}(\mu_{us})\alpha_{s}(\mu) \frac{C_{F}}{N_{c}^{2}}  
\nonumber \\ & 
 \int_{0}^1 dx\int_0^1  dy \left(
  \frac{\theta(\bar{x}<\eta) \theta(y<\eta)}{(y k+\bar{x} p)^{2}}D_{s}^{\alpha\beta}(x,y)
  +\frac{\theta(x<\eta) \theta(\bar{y}<\eta) }{(\bar{y}k+x p)^{2}}D_{q}^{\alpha\beta}(x,y)
  \right)
\nonumber  \\ &
\times \sqrt{N_{c}} \sqrt{M_\chi} \sqrt{\frac{3}{4\pi}}  \int \frac{d^{3}\vec{\Delta}}{(2\pi)^3}~\tilde{R}_{21}(\Delta)\frac{1}{4}
\text{tr}\left[  \Lambda_{J}(1-\Dsl{\omega})\gamma_{\bot\alpha}(1+\Dsl{\omega})\right] D_{Q}^{\beta}(E,\Delta_{\top}).
\label{usoft-d}
\end{align}
The second line of Eq.(\ref{usoft-d})  describes the  subdiagram with light quarks and ultrasoft gluon propagators (which gives expressions in the denominators).  The  $\theta$ functions restrict the integrations regions over the  quark collinear fractions  and  cut-off $\eta$ must be understood as UV-regulator.   The functions $D_s$  and $D_q$  consist of contributions of the light-quark vertices  and  DAs. A calculation of these
 contributions is the same as for the  colour-singlet case  but  in addition one has to  expand  the  integrands with respect to small fractions  in the regions $y\sim \bar{x}\sim v^2$ or $x\sim \bar{y}\sim v^2$ as it is indicated by the appropriate  $\theta$-functions.  

The third line in  Eq.(\ref{usoft-d}) describes the heavy quark  subdiagram.   The relative heavy quark momentum  $\Delta_\top$ is  of order $m_cv$, in what follow we  assume that $\Delta\equiv |\vec{\Delta}|$.  The momentum space  radial wave function of $P$-wave state  reads
\begin{equation}
\tilde{R}_{21}(\Delta)=iR_{21}^{\prime}(0)\frac{16\pi\gamma_{B}\Delta}{(\Delta^{2}+\gamma_{B}^{2}/4)^{3}},~\ \ \ \gamma_{B}=\frac{1}{2}%
m_{c}\alpha_{s}C_{F} .
\end{equation}%
where $R_{21}^{\prime}(0)$ is the derivative of the  position radial wave function at the origin.  The trace  over Dirac indices  includes the projectors  on $P$-wave state 
\begin{align}
\Lambda_{1} &  =\frac{1}{2\sqrt{2}} \frac
{\Delta_{\top}^\rho}{\Delta} \left[\gamma_\rho,\Dsl{\varepsilon}_{\chi}\right]  \gamma_{5},\,
\Lambda_{2}=-\varepsilon_{\chi}^{\mu\nu}\frac{\Delta_{\top\mu
}}{\Delta}\gamma_{\top\nu}.
\end{align}%
The factor  $\gamma_{\bot\alpha}$ originates from  the vertex of the octet $S$-wave operator in Eq.(\ref{A8hard}). A simple calculation yields
\bea
\frac{1}{4} \text{tr}\left[  \Lambda_{J}(1-\Dsl{\omega})\gamma_{\bot\alpha}(1+\Dsl{\omega})\right] =
-\frac{\Delta_\top\cdot (n-\bar{n})}{\Delta} \frac{1}2 
\left(\delta_{J1} {\sqrt{2}} i\varepsilon^\perp_{\rho \alpha  } \epsilon^\rho_{\chi}+\delta_{J2}2(\epsilon_{\chi})_{\alpha_\bot\beta}\bar{n}^\beta\right).
\label{trL}
\eea
The expression in the brackets  in Eq.(\ref{trL})  gives the dependence on the total momentum $J$ in the amplitude (\ref{usoft-d}). 
 The function  $D_{Q}^{\beta}(E,\Delta_{\top})$  in Eq.(\ref{usoft-d}) is described by the  chromoelectric  vertex  generated from the pNRQCD interaction Lagrangian 
\bea
\mathcal{L}_{int}(x)=-g \psi^\dag_\omega(x) \vec {x}\cdot\vec{E}(t)\psi_\omega(x)-g \chi^\dag_\omega (x)\vec {x}\cdot\vec{E}(t)\chi_\omega(x),
\eea
 and by octet Coulomb Green function $G_{C}$
\bea
D_{Q}^{\beta}(E,\Delta_{\top})= (\omega^\beta k^\lambda_g-g_\top^{\beta\lambda}(\omega k_g)) \frac{\partial}{\partial \Delta_\top^\lambda}
\int \frac{d^3\vec{\Delta}'}{(2\pi)^3}  G_{C}(\Delta_\top,\Delta'_\top; E-(\omega k_g) ),
\eea
where $k_g$  denotes the outgoing ultrasoft  gluon momentum ( remind that  $k_g=y k+\bar{x} p$ for the  $D^{\alpha\beta}_s$ and  $k_g=\bar{y} k+x p$ for $D^{\alpha\beta}_q$ ). 
 The expression for the  Coulomb Green function  is obtained by summation of the ladder diagrams with the colour-octet potential insertions. The resulting expression  is  quite complicated
 \bea
 G_{(8)}(\Delta_\top,\Delta'_\top; E )=-\frac{(2\pi)^3\delta^{(3)}(\vec{\Delta}-\vec{\Delta}')}{E+\Delta^2_\top/m_c}+\frac{g^2}{2N_c}\frac{1}{E+\Delta^2_\top/m_c}\frac{1}{(\vec\Delta-\vec\Delta')^2}\frac{1}{E+\Delta'^2_\top/m_c}+\mathcal{O}(g^4).
 \label{def:G8}
 \eea
  The full expression can be found in Refs.\cite{Schwinger:1964zzb, Beneke:2013jia}.  In the various calculations, see e.g. Refs.\cite{Beneke:2008pi, Beneke:2007pj,Beneke:2008cr,Brambilla:2011sg},   it has been observed that  the dominant  numerical   impact is provided by the  relatively simple  first term in Eq.(\ref{def:G8}), while the remnant higher order contributions are suppressed by  the factor $1/(2N_c)$ for each colour-octet exchange.  For our purpose  it is also enough to consider  an approximation which is given by no-gluon exchange leading term in Eq.(\ref{def:G8}).  This gives
\bea
D_{Q}^{\beta}(E,\Delta_{\top})\simeq  (\omega^\beta k^\lambda_g-g_\top^{\beta\lambda}(\omega k_g))\frac{2\Delta^\lambda_{\top}}{m_c}\frac{1}{\left[  E-(\omega k_{g})+\Delta_{\top
}^{2}/m_c+i\varepsilon\right]^{2}},
\label{DQ}
\eea
Substituting (\ref{trL}) and (\ref{DQ}) into Eq.(\ref{usoft-d}), and using rotation invariance in order  to  reduce $\Delta^{\rho}_\top\Delta^{\lambda}_{\top}\to - \Delta^2 g^{\rho\lambda}_\top/3$  one obtains 
\begin{align}
\left.\mathcal{M}_{\chi_{cJ}\to K^*\bar{K}}^{(8)} \right |_{\text{Fig.}\ref{fig_pNRQSD-diag}} & =
\left(\delta_{J1} {\sqrt{2}} i\varepsilon^\perp_{\rho \alpha  } \epsilon^\rho_{\chi}+\delta_{J2}2(\epsilon_{\chi})_{\alpha\rho}\bar{n}^\rho\right)
\frac{-2 \pi^{2} }{m_c^{2}}\alpha_{s}(\mu_{us})\alpha_{s}(\mu) \frac{C_{F}}{N_{c}^{2}}  
\nonumber \\ & 
 \int_{0}^1 dx\int_0^1  dy \left(
  \frac{\theta(\bar{x}<\eta) \theta(y<\eta)}{(y k+\bar{x} p)^{2}}D_{s}^{\alpha\beta}(x,y)
  +\frac{\theta(x<\eta) \theta(\bar{y}<\eta) }{(\bar{y}k+x p)^{2}}D_{q}^{\alpha\beta}(x,y)
  \right)
\nonumber  \\ &
\times \sqrt{N_{c}} \sqrt{M_\chi} \sqrt{\frac{3}{4\pi}} \frac{1}{m_c} \int \frac{d^{3}\vec{\Delta}}{(2\pi)^3}~\tilde{R}_{21}(\Delta)\frac{\Delta}{3}  
\frac{ \bar{n}^{\beta}(n k_{s})-n^{\beta}(\bar{n}k_{s}) }{\left[
E-(\omega k_{g})+\Delta_{\top}^{2}/m_c+i\varepsilon\right]  ^{2}}.
\label{usoft-d1}
\end{align}
Notice that  the total momentum  $J$  only  enters  in   expression in the brackets  in the first line.  The Dirac traces  in $D^{\alpha\beta}_{s,q}$  allows one to conclude  that the  expression in Eq.(\ref{usoft-d1})  can be presented  in the following form 
\bea
\left.\mathcal{M}_{\chi_{cJ}\to K^*\bar{K}}^{(8)} \right |_{\text{Fig.}\ref{fig_pNRQSD-diag}}& =
\left\{ 
\delta_{J1} {\sqrt{2}} i\varepsilon^\perp_{\rho \alpha  } \epsilon^\rho_{\chi}+\delta_{J2}2(\epsilon_{\chi})_{\alpha\beta}\bar{n}^\beta 
\right\}
(-i)\varepsilon_\perp^{\alpha\rho} (e^{*}_V)_\rho m_c \, J_{us}
\\
&=\left\{\delta_{J1} ( \epsilon_{\chi} e_V^*)-\delta_{J2} (\epsilon_{\chi})_{\alpha\beta}\bar{n}^\beta   i\varepsilon_\perp^{\alpha\rho} (e^{*}_V)_\rho    \right\}m_c  (-1)^J 2^{J/2} J_{us},
\eea
where $J_{us}$ is the  universal convolution integral.  Comparing  the last equation  with the definitions of the scalar amplitudes in Eqs.(\ref{M1:def}) and (\ref{M2:def}) one obtains
\begin{equation}
\mathcal{A}_{Jc}^{\bot (8)}=\left(  -1\right)  ^{J}2^{J/2} J_{us}.
\label{AJC}
\end{equation}
This result  already gives the relation in  Eq.(\ref{SSrel}).  In order to obtain Eq.(\ref{AJC}) we used the approximate expression for the  Coulomb Green function but the given derivation can also  be  extended to the case of exact colour-octet propagator.  

For the ultrasoft integral we  obtain 
\begin{equation}
J_{us}= \frac{ i\left\langle \mathcal{O}(^{3}P_{J})\right\rangle }{m_c^{3}}
\alpha_{s}(\mu) \alpha_{s}(\mu_{us})
\frac{\pi^{2}}{N_{c}^{2}}C_{F}  \left(\frac{f_{V}^{\bot}f_{P}\mu_{P}}{m_c^3} J_{us}[P_{3}V_{2}]+\frac{f_{P}f_{V}m_{V}}{m_c^3} J_{us}[P_{2}V_{3}]\right),
\end{equation}
with 
\begin{align}
J_{us}[P_{3}V_{2}]  =
m_c^2 \mathcal{N}\int \frac{ d^{3}\vec{\Delta}}{(2\pi)^3}~\tilde{R}_{21}(\Delta)\Delta 
\int_{1-\eta}^{1}dx\int_{0}^{\eta}dy\frac{ 
\bar{\phi}_{2V}^{\bot\prime}(1)\Delta\phi^{p}_{3P}%
(y\sim 0)+\Delta\phi_{2V}^{\bot\prime}(1)\bar{\phi
}^{p}_{3P}(y\sim 0)  }{\left[  E-m_c(y+\bar{x})+\Delta_{\top}%
^{2}/m_c+i\varepsilon\right]  ^{2}},
\label{JusP3V2}%
\end{align}
\begin{align}
J_{us}[P_{2}V_{3}]   =
\frac{m_c^2}2 \mathcal{N} \int \frac{ d^{3}\vec{\Delta}}{(2\pi)^3}\tilde{R}_{21}(\Delta)\Delta 
 \int_{1-\eta}^{1}dx\int_{0}^{\eta}dy \frac{\bar{\phi}_{2P} ^{\prime}(0)\left(  I[\Delta\Omega]-\Delta\Omega(1)\ln\bar
{x}\right)  + \Delta\phi_{2P}^{\prime}(0)I[\bar{\Omega}%
]}{\left[  E-m_c(y+\bar{x})+\Delta_{\top}^{2}/m_c+i\varepsilon\right]  ^{2}}.
\label{JusP2V3}
\end{align}
where  normalisation factor $\mathcal{N}$  is defined as 
\bea
\mathcal{N}=\frac13\frac{1}{iR'_{21}(0)} ,
\quad   \mathcal{N} \int \frac{ d^{3}\vec{\Delta}}{(2\pi)^3}~\tilde{R}_{21}(\Delta)\Delta=1.
\label{def:N}
\eea
  In Eq.(\ref{JusP2V3}) we  also defined the short notation
\bea
I[f]=\int_0^1du\frac{f(u)-f(1)}{1-u}.
\eea
Remind that  functions $\bar{f}$ and $\Delta f$ denote symmetric and antisymmetric  components, see Eq.(\ref{def:symasym}).

The ultrasoft integrals in Eqs.(\ref{JusP3V2}) and (\ref{JusP2V3}) are divergent if one takes  UV cut-off  $\eta\to \infty$. In order to compute these integrals  we must use the same regularisation  as for the colour-singlet case. Therefore we introduce  the analytical regularisation  substituting
\bea
\frac{1}{\left[  E-m_c(y+\bar{x})+\Delta_{\top}^{2}/m_c+i\varepsilon\right]  ^{2}}\to \frac{\nu^{2\varepsilon}}{\left[  E-m_c(y+\bar{x})+\Delta_{\top}^{2}/m_c+i\varepsilon\right]  ^{2+\varepsilon}}
\label{def:reg}
\eea 
and then take the limit $\eta\to \infty$.  Remind, that in  Eq.(\ref{def:reg}) we assume  $y\sim \bar{x}\sim v^2$ and therefore  all terms in the denominator are of order $mv^2$.  

Consider as example  the following term  from Eq.(\ref{JusP3V2}) 
\bea
J_{us1}[P_{2}V_{3}] =m_c^2 \mathcal{N}\int \frac{ d^{3}\vec{\Delta}}{(2\pi)^3}~\tilde{R}_{21}(\Delta)\Delta 
\int_{-\infty}^{1}dx\int_{0}^{\infty}dy\frac{ \nu^{2\varepsilon}
 \bar{\phi}_{2V}^{\bot\prime}(1)\Delta\phi^{p}_{3P}
(y\sim 0)}{\left[  E-m_c(y+\bar{x})+\Delta_{\top}
^{2}/m_c+i\varepsilon\right]  ^{2+\varepsilon}}.
\label{def:Ius}
\eea
  Performing  expansion of the integrand in the UV-region   $y\sim \bar{x}\to \infty$ one  can see the overlap  with the  singular collinear integral  in Eq.(\ref{Ising}). 
 The regularised integral in Eq.(\ref{def:Ius})  can be easily computed  separating  the divergent part in the following way
\bea
 && J_{us1}[P_{3}V_{2}] =m_c^2 \mathcal{N}\int \frac{ d^{3}\vec{\Delta}}{(2\pi)^3}~\tilde{R}_{21}(\Delta)\Delta   \int_{0}^{\infty}dx\int_{0}^{\infty}dy 
 \left(   \bar{\phi}_{2V}^{\bot}\right)  ^{\prime}(1)  \Delta\phi^{p}_{3P}(y\sim 0) 
\nonumber  \\ &&
\times \left(
\frac{1 }
{\left[  E-m_c(y+x)+\Delta_{\top}^{2}/m_c+i\varepsilon\right]^{2} }
- 
\frac{ 1}
{\left[  E-m_c(y+x)+i\varepsilon\right]^{2} }\right)
\nonumber \\ &&
+m^2 \int_{0}^{\infty}dx\int_{0}^{\infty}dy
\frac{\nu^{2\varepsilon}\bar{\phi}_{2V}^{\bot\prime}(1)  \Delta\phi^{p}_{3P}(y\sim 0)}
{\left[  E-m_c(y+x)+i\varepsilon\right]^{2+\varepsilon} },
\label{sep:Jus1}
\eea
where we used  relation (\ref{def:N}).
The first integral in (\ref{sep:Jus1}) is finite and therefore  the regularisation in this case can be omitted. Then the  integrals over the collinear fractions can be easily computed and one finds ($t=\nu^2/m_c^2$)
  \bea
 &&J_{us1}[P_{3}V_{2}] = \bar{\phi}_{2V}^{\bot \prime}(1)\frac{3}{2}\rho_{-}^{K}
 \left\{  \frac{\beta}{\varepsilon^{2}}+\frac{1}{\varepsilon}(\beta\ln t+\alpha-\beta)+
 \frac{\beta}{2}\ln^{2}t+(\beta-\alpha)(1-\ln t)-\beta\frac{\pi^{2}}{6}\right. 
\nonumber \\  && \left.
 -\mathcal{N}\int  \frac{ d^{3}\vec{\Delta}}{(2\pi)^3}~
 \tilde{R}_{21}(\Delta)\Delta \left(  \frac{1}{2}\beta \ln^{2}\left[  \Delta^{2}/m_c^{2}%
-E/m_c-i0\right]  +\alpha \ln\left[  \Delta^{2}/m^{2}-E/m_c-i0\right] 
\right)  \right\} ,
\label{res:Jus1}
 \eea
 with $\alpha$ and $\beta$ defined in Eq.(\ref{log0}).
 Comparing this result with the expression in Eq.(\ref{res:Jc1div})  one can see that the poles and  logarithms $\ln t$ cancel in the sum $J^{(J)}_{c1}[P_{2}V_{3}] +(-1)^J J_{us1}[P_{2}V_{3}]$.  The computation of the other integrals in Eqs.(\ref{JusP3V2}) and (\ref{JusP2V3}) is  similar,  we have  checked that the poles and $\mu$-dependence  also cancel in the sum of singlet and octet amplitudes.  This  demonstrates that  our matching  is consistent  and  the endpoint IR singularities in the colour-singlet amplitude can be absorbed into colour-octet  contribution.  Therefore at least to a given accuracy the sum of the singlet and octet amplitudes describes the physical amplitude consistently.  
  
 The  octet integrals in Eqs.(\ref{JusP3V2}) and (\ref{JusP2V3}) include the derivatives of  DAs  and  integrals with the radial wave function as in Eq.(\ref{res:Jus1}).  Such integrals generate imaginary part which appears from the region where $\Delta^{2}/m_c< E$   and can be associated with the cut  of diagrams as shown in Fig.\ref{fig_pNRQSD-diag}.  The imaginary part  is a  direct consequence of the intermediate colour-octet state. The expression (\ref{res:Jus1}) also  demonstrates  that octet contribution is sensitive to the shape of the radial  wave function $\tilde{R}_{21}$ while  the  singlet amplitude depends only from the wave function at the origin.  This is qualitative difference between the two terms and it is interesting to study, at least qualitatively,  how  this point can affect  a description of quarkonium decays.

 Consider now  the diagrams in Fig.~\ref{fig_softoverlap}.  Their computation  can be  done within the same technique as described above. The coupling of the ultrasoft gluon to the hard-collinear quarks is described by the leading-order SCET interactions $ \bar{\psi}_n(x) g(n\cdot A_{us}((x\bar{n})n/2))\Dsl{\bar{n}}/2\psi_n(x)$ and similarly for the $\bar{n}$ light-cone sector. 
 This again yields the result  (\ref{AJC}) for the corresponding amplitudes
\begin{equation}
\mathcal{A}_{Js}^{\bot (8)}=\left(  -1\right)  ^{J}2^{J/2} \tilde{J}_{us}.
\end{equation}
    The ultrasoft integral  in this case  reads
\begin{align}
\tilde{J}_{us} =&\frac{i\left\langle \mathcal{O}(^{3}%
P_{0})\right\rangle }{m_c^{3}}~  \alpha_{s}(\mu_{h})\alpha
_{s}(\mu_{us})\frac{C_{F}}{2N_{c}}~\left(  f_{PV}^{~s}-f_{PV}^{~q}\right) 
\mathcal{N}\int d^{3}\vec{\Delta}\tilde{R}_{21}(\Delta)\Delta
\nonumber \\ &
\times \frac{1}{i\pi^{2}} \int d^Dl  
\frac{1}{\left[  l^{2}\right]  }\frac{ }{\left[  E-m_c(l\omega)+\Delta_{\top
}^{2}/m_c+i\varepsilon\right]  ^{2}}.
\end{align}
The integral over the ulrasoft momentum $l$ is UV-divergent and we use  dimensional regularisation as before,  with $d^Dl= e^{\varepsilon(\gamma_E+\ln\pi)}\mu^{2\varepsilon}_{us}\,  d^{4-2\varepsilon} l$.  Calculation of this  integral yields     
\begin{equation}
 \frac{1}{i\pi^{2} } \int d^Dl  
\frac{1}{\left[  l^{2}\right]  }\frac{ }{\left[  E-m_c(l\omega)+\Delta_{\top
}^{2}/m_c+i\varepsilon\right]  ^{2}} =-\frac{1}{\varepsilon}+\ln\frac{m_c^2}{\mu^2}+2\ln[2\left(  \Delta^{2}/m_c-E-i0\right)  /m_c].
\label{res:Jscet}
\end{equation}
Therefore  we  obtain 
\begin{align}
A_{J=2,s}^{\bot (8)}  & =2\tilde{J}_{us} =\frac{i\left\langle \mathcal{O}(^{3}%
P_{0})\right\rangle }{m_c^{3}}~ \alpha_{s}(\mu)\alpha
_{s}(\mu_{us})\frac{C_{F}}{N_{c}}~\left(  f_{PV}^{~s}-f_{PV}^{~q}\right) 
\nonumber  \\
& \times \left( 
-\frac{1}{\varepsilon}+\ln \frac{m_c^2}{\mu^2}+2\ln2+2\mathcal{N}\int \frac{d^{3}\vec{\Delta}}{(2\pi)^3}\tilde{R}_{21}(\Delta)\Delta \ln[\left(  \Delta^{2}/m_c-E-i0\right) /m_c]
 \right).
 \label{A2s8}
\end{align}
Comparing this expression with the hard contribution in (\ref{res:Cgg}) we observe  that poles in $1/\varepsilon$ and $\mu$-dependence  cancel in the sum $A_{J,s}^{\bot (0)}+A_{J,s}^{\bot (8)} $.  The soft-overlap amplitude (\ref{A2s8}) also has imaginary part  which is  generated by the cut shown in Fig.\ref{fig_softoverlap}. 

To summarise.  The  colour-octet amplitudes defined in Eq.(\ref{A8hard})  are given by the sum 
\begin{align}
\mathcal{A}^{\bot (8)}_J = \mathcal{A}^{\bot (8)}_{Jc} + \mathcal{A}^{\bot(8)}_{Js}. 
\end{align}
The total decay amplitudes are given by the sum of the singlet and octet amplitudes (\ref{Aperp: fact}), the singular terms cancel in this sum so that decay amplitude  is well defined. This cancellation  allows us conclude that  various  IR-singularities which have been observed  in the colour-singlet amplitudes  can be  absorbed into renormalisation of the  colour-octet matrix element (\ref{A8hard}).  This matrix element is sensitive to a long-distance behaviour of the  quarkonium wave function and have   imaginary part due to long distance  interactions.  Can one get any information about  the colour-octet contribution from the experimental data?  We try to study this question in the next section.

\section{Phenomenology}
\label{phen}
 In Sec.3  we obtained that  the colour-singlet amplitude $\mathcal{A}_{1}^{\Vert}$ provides a tiny  contribution and cannot describe the measured  branching ratio. Hence we can suppose  that the dominant effect  is provided by the  transverse amplitudes which  are given by the sum of the colour-singlet and colour-octet  terms.
Suppose that the largest numerical effect is provided by the colour-octet amplitudes $\mathcal{A}^{\perp(8)}_J$, i.e. $\mathcal{A}_{J}^{\perp(8)}\gg\mathcal{A}_{J}^{\perp(0)}$.  In the previous section it was  established  that these amplitude satisfy to Eq.(\ref{SSrel})
up to  relativistic corrections in velocity $v$.   Using this  relation  and Eqs.(\ref{Gamma1}) and (\ref{Gamma2}) one  obtains
\begin{equation}
R_{th}=\frac{\Gamma[\chi_{c2}\to \bar{K}^0K^{*0}+c.c.]}{\Gamma[\chi_{c1}\to \bar{K}^0K^{*0}+c.c.]}\simeq\left(  1-\frac{m_{P}^{2}}{k_{0}^{2}}\right)  \left(  1-\frac{m_{P}%
^{2}m_{V}^{2}}{(kp)^{2}}\right)  \frac{1}{5}\frac{3}{2}\frac{\left\vert
\mathcal{A}_{2}^{(8)}\right\vert ^{2}}{\left\vert \mathcal{A}_{1}%
^{(8)}\right\vert ^{2}}=0.55,
\end{equation}
This estimate  includes  contribution  from the model dependent power suppressed coefficient  which yields
\begin{equation}
\left(  1-\frac{m_{P}^{2}}{k_{0}^{2}}\right)  \left(  1-\frac{m_{P}
^{2}m_{V}^{2}}{(kp)^{2}}\right)=0.91.
\end{equation}
Using the data for neutral mesons $\bar{K}^0$ and $K^{*0}$ from Table~\ref{dataXcJ}  one finds
\begin{equation}
R_{\text{exp}}=\frac{\text{Br}[\chi_{c2}\rightarrow \bar{K}K^*+c.c.]}{\text{Br}[\chi
_{c1}\rightarrow \bar{K}K^*+c.c.]}\frac{\Gamma_{tot}[\chi_{c2}]}{\Gamma_{tot}[\chi_{c1}]}=0.30\pm0.13,
\end{equation}
where we used $\Gamma_{tot}[\chi_{c1}]=0.84$~MeV and $\Gamma_{tot}[\chi_{c2}]=1.93$~MeV  \cite{Patrignani:2016xqp}.
The  difference  of about factor two between the   values $R_{\text{th}}$ and $R_{\text{exp}}$ allows one to suppose  that  effect from the colour-singlet  contribution  is not negligible and could help to improve the description.  For simplicity we consider  only  branching fractions  of  the neutral mesons. The decay amplitudes of the neutral and charged mesons must be the same due to $SU(2)$ flavour symmetry and data support this conclusion.  Therefore a consideration of the decays of charged mesons  provides the  similar  results.

 The  colour-singlet  HQSS breaking relations have been obtained in Eqs.(\ref{def:DA0}) and (\ref{res:A0s}).  Using them we can relate the decay  amplitudes as 
\begin{align}
\mathcal{A}_{1}^{\bot}  =\Delta\mathcal{A}_{c}^{\bot(0)}+\Delta\mathcal{A}_{s}^{\bot(0)}-\frac{1}{\sqrt{2}}\mathcal{A}_{2}^{\bot}.
\label{res:A1}
\end{align}

The absolute value $|\mathcal{A}_{2}^{\bot}|$ can be estimated from the  width $\Gamma[\chi_{c2}\rightarrow \bar{K}^0K^{*0}+c.c.]$ that gives
\begin{equation}
\left\vert \mathcal{A}_{2}^{\bot}\right\vert =\left(  7.0\pm1.5\right)
\times10^{-3}.
\label{A2perpnum}
\end{equation}
The result for absolute   value  $|\mathcal{A}_{1}^{\bot}|$  which can be obtained from  Eq.(\ref{res:A1})  depends on the unknown imaginary phase of amplitude  $\mathcal{A}_{2}^{\bot}$
\begin{equation}
~\ \mathcal{A}_{2}^{\bot}=\left\vert \mathcal{A}_{2}^{\bot}\right\vert
e^{i\delta},
\end{equation} 
and on  the unknown difference of the SCET amplitudes $f_{PV}^{s}-f_{PV}^{q}$ in $\Delta\mathcal{A}_{s}^{\bot(0)}$, see Eq.(\ref{A2s8}).  We accept these quantities as unknown parameters.  Let us rewrite  the soft-overlap combination  as 
\begin{equation}
f_{PV}^{s}-f_{PV}^{q}=\frac{f_{P}f^{\Vert}_{V}m_{V}}{m_c^{3}}\Delta f,
\end{equation}
where   factor  ${f_{P}f^{\Vert}_{V}m_{V}}/{m_c^{3}}$  introduces a ``natural" scale. In the following  we assume  that  parameter  $\Delta f$ is real.  We can not provide a rigorous arguments about a suppression of the imaginary part of  $\Delta f$ and therefore accept  this simplification as reliable assumption.    
 
 In order to get numerical estimates we use the following non-perturbative input.  The  models of  $K$-meson DAs, quark masses  and  numerical estimates for  NRQCD matrix elements  are described in Appendix~A.  Calculating  symmetry breaking corrections $\Delta\mathcal{A}^{\bot (0)}_{Jc,s}$  we use  $n_f=4$, $m_c=1.5$~GeV and  set the value of renormalisation scale $\mu^2=2m_c^2$  that gives $\alpha_s(2m_c^2)=0.29$.   We  also apply the leading logarithmic evolution for the  parameters of DAs.   
 
 The  expression for  $\Delta\mathcal{A}_{c}^{\bot(0)}$ is  described in Eqs.(\ref{def:DA0})-(\ref{res:sumJcP3V2}) and using the numerical values of the DA parameters  we obtain 
\begin{equation}
\Delta\mathcal{A}_{c}^{\bot(0)}=(-1.56\pm0.19)\times 10^{-3},
\label{DAcnum}
\end{equation}
where the errors give the  uncertainty from the variation of values of the DA parameters.   Both
contributions in Eq.(\ref{def:DA0}) are negative, the largest numerical
impact  is provided  by the terms proportional to $SU(3)$-breaking parameters  $\rho_{-}^{K}$ and
$\lambda_{s}^{-}$, see definitions in Eqs.(\ref{def:phip3P}) and (\ref{def:lambda}).   The chiral enhanced contribution associated with the projection $P_{3}V_{2}$  is   about factor two larger than the contribution from the  $P_{2}V_{3}$ projection. Comparing results for  amplitude $|\mathcal{A}_{2}^{\bot}|$ in Eq.(\ref{A2perpnum}) and for  $\Delta\mathcal{A}_{c}^{\bot(0)}$ in Eq.(\ref{DAcnum})  one finds  that  the value of the symmetry breaking corrections are few times smaller.  

 \begin{figure}[h]%
\centering
\includegraphics[width=6.0in]%
{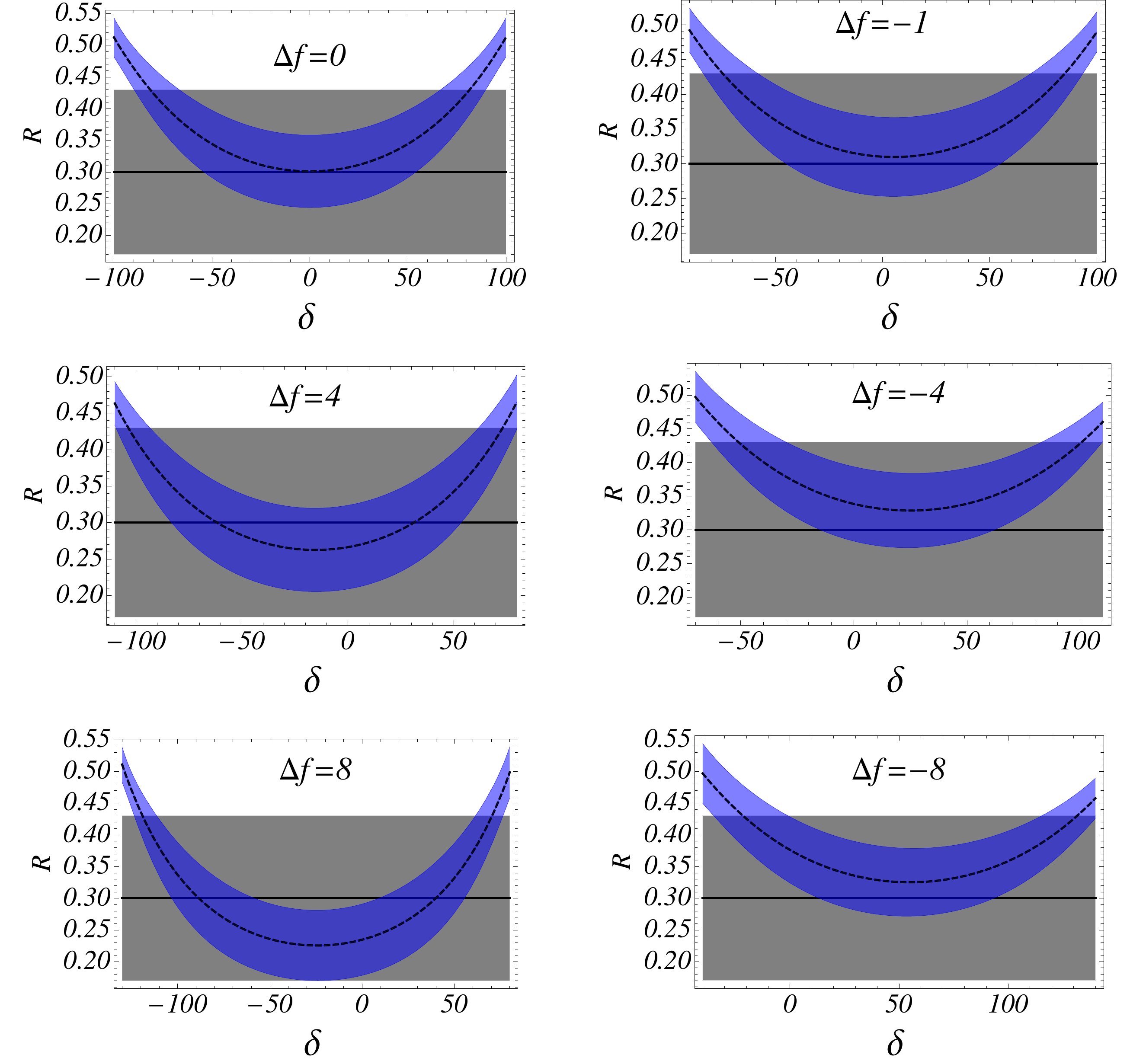}%
\caption{Ratio $R_{\text{th}}$ (dashed) as a function of  angle $\delta$ (in degrees)  for the  different fixed values of parameter $\Delta f$. The experimental value $R_{\text{exp}}$ is shown by solid line. The blue and gray shaded areas  show  theoretical and experimental uncertainties, respectively. }%
\label{fig_plot-rf}%
\end{figure}
For the symmetry breaking soft-overlap contribution (\ref{res:A0s})  we obtain 
\begin{equation}
\Delta\mathcal{A}_{s}^{\bot(0)}=\left(
-0.10~+0.13i\right)  \Delta f\times10^{-3},
\end{equation}
where  $\Delta f$ is unknown  parameter. In the following  we suppose   that the colour-singlet soft-overlap contribution is smaller or of the same order as $\Delta\mathcal{A}_{s}^{\bot(0)}$
\begin{equation}
|\Delta\mathcal{A}_{s}^{\bot(0)}(\Delta f)|\lesssim|\Delta \mathcal{A}_{c}^{\bot(0)}|,
\label{Dfsmall}
\end{equation}
that implies  $|\Delta f|\lesssim10$.  In this case  one obtains, for instance, 
\begin{align}%
\Delta\mathcal{A}_{s}^{\bot(0)}(\Delta f=4)=(-0.41+0.54i)\times10^{-3} ,\quad \Delta\mathcal{A}_{s\bot}^{(0)}(\Delta f=8)=(-0.81+1.1i) \times10^{-3}
\end{align}

Numerical  estimates of $R_{\text{th}}$ in comparison with the $R_{\text{exp}}$  are shown in Fig.\ref{fig_plot-rf}. The theoretical error band (blue shaded area) corresponds to variation of the DA parameters and value $| \mathcal{A}_{2}^{\bot}|$ according to result in Eq(\ref{A2perpnum}).  We see that for each value of $\Delta f$ we have sufficiently large interval  for the phase $\delta$ which allows to describe the ratio $R_{\text{exp}}$  within the error bars.   The largest numerical effect from the symmetry breaking corrections  is provided by the interference with large amplitude  $\mathcal{A}_{2}^{\bot}$. From Fig.\ref{fig_plot-rf}   we  conclude that  reliable  description of the data for the branching fractions can only  be done taking into account both colour-octet and colour-singlet  amplitudes. 

In the previous sections  it was  shown that IR-singularities which appear in the convolution integrals of the colour-singlet amplitudes can be absorbed into the colour-octet  amplitudes. Therefore one can define a regular colour-singlet contribution  by  subtraction of  the  IR-poles.   Using such  definition of  the colour-singlet amplitudes  one can try to estimate the value  of  colour-octet amplitude from the phenomenological value $\mathcal{A}_2^\perp$  obtained in Eq.(\ref{A2perpnum}). 

\begin{figure}[h]%
\centering
\includegraphics[
width=7.0in
]%
{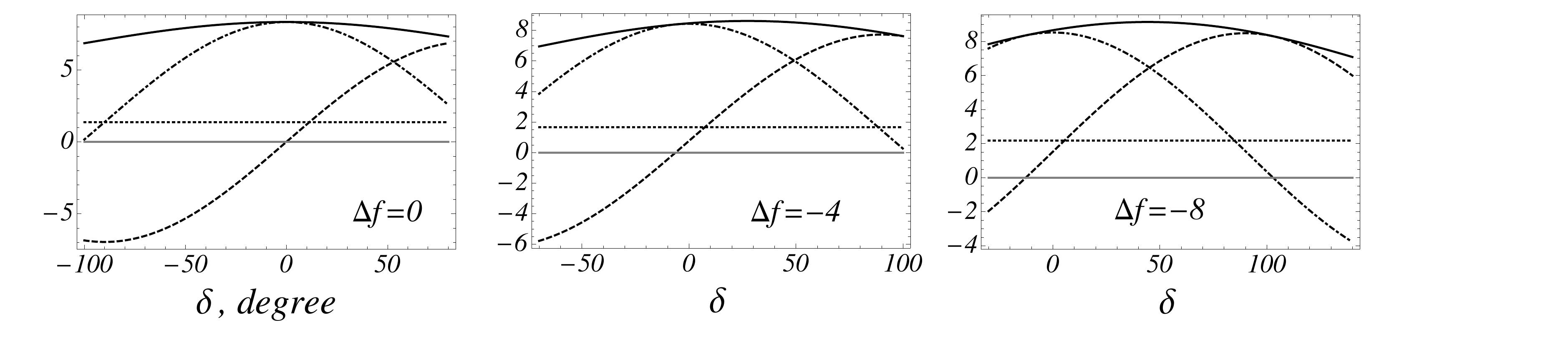}%
\caption{The  colour-octet amplitude $A_{2}^{\bot(8)}$ in units
$10^{-3}$ as a functions of angle $\delta$ at fixed $\Delta f$. The solid,
dot-dashed and dashed lines correspond to  absolute value, real and
imaginary parts, respectively. The dotted line shows the absolute value of the
colour-singlet amplitude $|A_{2}^{\bot(0)}|$. }%
\label{fig_plot-a28}%
\end{figure}

The colour-singlet amplitude is given by  sum of the collinear $\mathcal{A}_{2c}^{\perp(0)}$  and soft-overlap $\mathcal{A}_{2s}^{\perp(0)}$  contributions given in  Eqs.(\ref{res:A0Jbot}) and (\ref{def:A0Js}), respectively.  The analytical expressions for the finite part of the collinear integrals are presented  in Appendix B.   Both amplitudes depends on the  factorisation scales $\mu$ or $\nu$ which are set to $m_c$.  Such scale setting  removes  the logarithms  $\ln m_c/\mu$  which must cancel in the sum of singlet and octet amplitudes.   The  values for parameters  $\Delta f$ and $\delta$  are chosen according to results in  Fig.~\ref{fig_plot-rf}.  This allows us to  obtain numerical estimates for the real and the imaginary parts  of the  colour-octet amplitude $\mathcal{A}_{2}^{\perp(8)}$.  The  obtained results are shown  in Fig.~\ref{fig_plot-a28}. 
  We see that  the absolute  value of the colour-octet amplitude is always few times larger  than the colour-singlet one. But the values of the real and imaginary parts of the octet amplitude strongly  depend on the phase $\delta$.  
  
  If  the soft-overlap amplitude  in Eq.(\ref{Dfsmall}) is underestimated then the value of the colour-singlet amplitude can be larger.  But even if we take  $\Delta f=20$ the colour-octet corrections remain sufficiently  large and important.  Therefore at least qualitatively  we definitely can conclude that colour-octet  mechanism plays very important role in the description of   $\chi_{Jc}\to \bar{K} K^*$ decays. 

\section{Discussion}
\label{conc}

Motivated by existing  experimental data we discuss  a description of  decay amplitudes $\chi_{cJ}\to \bar{K} K^*$  within the effective field theory framework. We find that  the leading-order amplitude, which  describes  decay $\chi_{c1}\to \bar{K} K^*_\Vert$,  is described  by  the colour-singlet operator but the  corresponding  contribution gives only about  few percents  of  the measured branching ratio. We expect that the dominant effect is given by the subleading  amplitudes which describe decays into  transversely polarised vector meson  $\chi_{cJ\perp}\to \bar{K} K_\perp^*$.  The colour-singlet contributions for   these  amplitudes involve  combinations  of  twist-2 and twist-3 collinear matrix elements  as required by helicity conservation.  In order to simplify our consideration  we  perform our calculations  in  the  Wandzura-Wilczek approximation  neglecting  the  twist-3  quark-gluon  matrix elements.  
 The computed colour-singlet amplitudes include  the  collinear convolution integrals which have  infrared endpoint divergences.  The  structure  of these singularities clearly indicates  the mixing  with  a colour-octet operator.  The  corresponding colour-octet matrix element  has been  studied in the Coulomb limit  using the pNRQCD  framework.  We obtain that  UV-singularities  of the octet contribution  exactly reproduce  the IR-singularities  of the  colour-singlet one  and therefore  these  IR-divergencies  can be absorbed into the renormalisation of the colour-octet matrix element.  Hence a  consistent description  of the  decay amplitudes $\chi_{cJ\perp}\to \bar{K} K_\perp^*$ is only given by  the sum of  colour-singlet and colour-octet matrix elements. The effective field theory  calculations  also allow us to establish  that  the colour-octet amplitude has an imaginary part which  is generated by the cut of the  intermediate state  with  the bound heavy quark-antiquark  in the octet configuration. 

The  heavy quark spin symmetry  allows us to establish  a relation between the colour-octet matrix elements for vector $J=1$ and tensor $J=2$ states.  This makes it possible to do  a computation of the spin symmetry breaking corrections which are free from  the  IR-divergencies.  We compute these corrections  using  the physical subtraction scheme.   We also include in our description a contribution of an unknown long distance  matrix element  describing soft-overlap configuration of the final mesons.   Making  various assumptions  about the value of this matrix element  we obtain a reliable description of  the branching fractions.  We conclude  that  the colour-octet contribution  must be   few times larger  then the colour-singlet one.

The uncertainties in our  consideration  can be  considerably  reduced  if one provides  an estimate for  the unknown  soft-overlap matrix element.  Potentially,  the model independent  information about this quantity can be obtained from the cross section of $\gamma\gamma\to \bar{K}K^*$  process  in the kinematical region where  $s\sim-t\sim -u\gg \Lambda^2_{QCD}$. One can also compute the contributions with the  twist-3 quark-gluon distribution amplitudes which have been discarded in this paper. We suppose that such contributions will improve the theoretical description but they will not change the qualitative conclusions of this work.

\section*{Aknowlegements}
I am grateful to M.~Vanderhaeghen for attracting my attention to work  \cite{BESIII:2016dda}  and for the discussions.  

\section{Appendix A. The long distance matrix elements}

The NRQCD long distance  matrix elements are defined as, see {\it e.g.} Ref.\cite{Bodwin:1994jh, Beneke:2008pi} 
\begin{equation}
\left\langle 0\right\vert \frac{1}{2\sqrt{2}}~\chi _{\omega }^{\dag }%
\overleftrightarrow{D}_{\top }^{\alpha }\left( \frac{-i}{2}\right) \left[
\gamma _{\top }^{\alpha },\gamma _{\top }^{\beta }\right] \gamma _{5}\psi
_{\omega } \left\vert \chi
_{c1}(\omega )\right\rangle =\epsilon _{\chi }^{\beta }i\left\langle 
\mathcal{O}(^{3}P_{1})\right\rangle ,  
\label{def:O3P1}
\end{equation}%
\begin{equation}
\left\langle 0\right\vert\chi _{\omega }^{\dag }\left( -\frac{i%
}{2}\right) \overleftrightarrow{D}_{\top }^{(\alpha }\gamma _{\top }^{\beta
)}\psi _{\omega }\left\vert
\chi _{c2}(\omega )\right\rangle =\epsilon _{\chi }^{\alpha \beta
}i\left\langle \mathcal{O}(^{3}P_{2})\right\rangle ,  \label{def:O3P2}
\end{equation}%
 The combination $(\alpha ,\beta )$ denotes the symmetrical traceless tensor These
operators are constructed from the quark $\psi _{\omega }$ and antiquark $%
\chi _{\omega }^{\dag }$ four-component spinor fields satisfying $\NEG%
{\omega}\psi _{\omega }=\psi _{\omega }$, $\NEG{\omega}\chi _{\omega }=-\chi
_{\omega }$.  The constants on the \textit{rhs }of Eqs.(\ref{def:O3P1}) and (%
\ref{def:O3P2}) are related to the value of the charmonium wave functions at
the origin. To leading order in small velocity $v$ they read \cite{Beneke:2008pi}
\begin{equation}
\left\langle \mathcal{O}(^{3}P_{J})\right\rangle \simeq \sqrt{2N_{c}}\sqrt{%
2M_{\chi_{cJ}}}\sqrt{\frac{3}{4\pi }}R_{21}^{\prime }(0),  
\label{<O3PJ>}
\end{equation}%
where $R_{21}^{\prime }(0)$ is the derivative of the quarkonium radial wave
function. The value of this parameter has been estimated  in the different potenial models, see e.g. Ref.\cite{Eichten:1995ch}. In this paper we use  the value  
computed for the Buchm\"uller-Tye potential
\begin{equation}
m_c=1.5\text{ GeV}, \, \, |R_{21}^{\prime }(0)|^{2}=0.75\, \text{GeV}^{5}.
\end{equation}

The leading twist $K$ and $K^*$ meson DAs has been studied in many publications, see e.g. \cite{Khodjamirian:2004ga,Braun:2004vf,Ball:2005vx,Braun:2006dg,Boyle:2006pw} and references there in.  The new   updates  of twist-2 and twist-3  DAs can be found in  Refs.\cite{Ball:2006wn,Ball:2007rt, Ball:2007zt}.  For a convenience of the reader,  we briefly describe definitions and models which are used in this paper.  

In the following we assume that direction $z_\mu$ is  light-like ($z^2=0$) and $[z,-z]$ denotes the appropriate Wilson line, see e.g. Ref.\cite{Ball:2006wn}.  For pseudoscalar state $\bar{K}(\bar{q}s)$ we need the  following matrix elements
\begin{equation}
\left\langle \bar{K}(k)\right\vert \bar{s}(z)[z,-z]\gamma^{\mu}\gamma_{5}q(-z)\left\vert
\right\rangle =-if_{K}k^{\mu}\int_0^1 dy~e^{i(2y-1)(kz)}\phi_{2\bar{K}}(y),
\label{phi2K}
\end{equation}%
\begin{equation}
\left\langle \bar{K}(k)\right\vert \bar{s}(z)[z,-z]i\gamma_{5}q(-z)\left\vert \right\rangle
=f_{K}\mu_{K}\int dy~e^{i(2y-1)(kz)}%
\phi_{3\bar{K}}^{p}(y),
\end{equation}%
\begin{equation}
\left\langle \bar{K}(k)\right\vert \bar{s}(z)[z,-z]\sigma_{\mu\nu}\gamma_{5} q(-z)\left\vert \right\rangle
=\frac{i}{3}f_{K}\mu_{K}\left(  k_{\mu}z_{\nu}-k_{\nu}z_{\mu}\right)  
\int_0^1 dy~e^{i(2y-1)(kz)}\phi_{3\bar{K}}^{\sigma}(y).
\end{equation}
Here 
\bea
\mu_{K}=m_{K}^{2}/(m_{s}+m_{q}),
\label{def:muK}
\eea
and    $m_{K}, m_s, m_q$ denote the masses of the $K$-meson, $s$- and  $q=u,d$ quarks,  $f_{K}$ is the decay constant.   The models for the corresponding  DAs  are given by the sum of the  few  first Gegenbauer moments  ($\bar{y}\equiv 1-y$)
\begin{equation}
\phi_{2\bar{K}}(y,\mu)=6y\bar{y}\left(  1+b_{1}(\mu)C_{1}^{3/2}(2y-1)+b_{2}(\mu)C_{2}^{3/2}(2y-1)\right)  ,
\label{mod:phiP}
\end{equation}
\begin{equation}
\phi_{3\bar{K}}^{p}(x,\mu)=1+\rho_{-}^{K}\frac{3}{2}(1+6b_{2}(\mu))\ln\frac{x}{\bar{x}}%
-\rho_{-}^{K}b_{2}(\mu)\frac{9}{2}\left\{  6~C_{1}^{1/2}(2x-1)+C_{3}^{1/2}%
(2x-1)\right\} ,
\label{def:phis3P}
\end{equation}%
\begin{equation}
\phi_{3\bar{K}}^{\sigma}(x,\mu)=6x\bar{x}\left\{  1-\rho_{-}^{K}b_{2}(\mu)\frac{15}{2}%
C_{1}^{3/2}(2x-1)\right\}  +\rho_{-}^{K}(1+6b_{2}(\mu))~9x\bar{x}~\ln\frac{x}%
{\bar{x}},
\label{def:phip3P}
\end{equation}
where $\rho_{\pm}^{K}=({m_{s}^{2}\pm m_{q}^{2}})/{m_{K}^{2}}$, in Eqs.(\ref{def:phis3P}) and (\ref{def:phip3P}) we neglect numerically  small terms  $\sim\rho_{\pm}^{K}b_{1}$.  Remind, that in this paper we do not consider   the three-particle quark-gluons operators therefore such term are also neglected in expressions in  Eqs.(\ref{def:phis3P}) and (\ref{def:phip3P}).   
 The evolution of the various DA parameters  are well known and explicit formulas can be found in the given references.  In our numerical calculations we use the following  values
\begin{equation}
\ m_{\bar{K}^0}=498\text{ MeV},~\ m_{s}(2\text{ GeV})=100 \text{ MeV},~\ m_{u}\simeq m_{d}\approx 0.
\end{equation}
\begin{equation}
f_{K}=0.160\text{ MeV},\, \,   b_{1}(1\text{ GeV})=0.06\pm0.03\,, \,  b_{2}(1\text{ GeV})=0.30\pm0.15.
\end{equation}

The required light-cone matrix elements for vector meson  $K^*(q\bar{s})$   read
\begin{equation}
\left\langle K^*(p,e^*)\right\vert  \bar{q}(z)[z,-z]\sigma_{\mu\nu}s(-z)\left\vert
0\right\rangle =-if^\perp_{K^*}m_{K^*} (e^*_\mu p_\nu- e^*_\nu p_\mu) 
\int_{0}^{1}dx e^{i(2x-1)(pz)}
\phi_{2K^*}^{\perp}  (x)  ,
\end{equation}%
\begin{equation}
\left\langle K^*(p,e^*)\right\vert  \bar{q}(z)[z,-z] \gamma_{\sigma}s(-z)\left\vert
0\right\rangle =-if^\Vert_{K^*}m_{K^*}\int_{0}^{1}dx e^{i(2x-1)(pz)}\left\{
p_{\sigma}\frac{\left(  e^{\ast}z\right)  }{(pz)}\phi_{2K^*}^{\Vert}%
(x)+ e_{\sigma_\bot}^{\ast} \phi_{3 K^*}^{\perp}(x)\right\}  ,
\end{equation}%
\begin{equation}
\left\langle  K^*(p,e^{\ast})\right\vert \bar{q}(z)[z,-z]\gamma_{\sigma}\gamma_{5}s(-z)\left\vert 0\right\rangle 
= f^\Vert_{K^*}m_{K^*} \frac{1}{2}i\varepsilon
_{\sigma\rho \mu\nu}e^{\ast\rho}p^\mu z^\nu \int_{0}^{1}dx e^{i(2x-1)(pz)} \psi^{\perp}_{3K^*}(x).
\end{equation}
The corresponding  models of DAs  read 
\begin{equation}
\phi_{2K^*}^{\Vert,\bot}(u)=6u\bar{u}(1+a_{1K^*}^{\Vert,\bot}C_{1}^{3/2}%
(2u-1)+a_{2K^*}^{\Vert,\bot}C_{1}^{3/2}(2u-1)),
\label{mod:phiIIV}
\end{equation}
\begin{equation}
\phi_{3K^*}^{\bot}(u)=\frac{1}{2}\int_{0}^{u}\frac{dv}{\bar{v}}\Omega
(v)+\frac{1}{2}\int_{u}^{1}\frac{dv}{v}\Omega(v)+\lambda_{s}^{+}\phi
_{2V}^{\bot}(u),\label{WWphiT}%
\end{equation}%
\begin{equation}
\psi_{3K^*}^{\bot}(u)=2\bar{u}\int_{0}^{u}\frac{dv}{\bar{v}}\Omega(v)+2u\int%
_{u}^{1}\frac{dv}{v}\Omega(v),
\end{equation}
with%
\begin{align}
\Omega(u) &  \simeq\phi_{2K^*}^{\Vert}(u)+\lambda_{s}^{+}\frac{1}{2}(2u-1)\frac{d}{du}\phi_{2K^*}^{\bot
}(u)+\lambda_{s}^{-} %
\frac{1}{2}\frac{d}{du}\phi_{2K^*}^{\bot}(u),
\label{def:Omega}\\
\lambda_{s}^{-}&=\frac{f_{K^*}^{\bot}}{f_{K^{\ast}}^{\Vert}}\frac{m_{s}-m_{q}}{m_{K^*}}, \,\,
\lambda_{s}^{+}= \frac{f_{K^*}^{\bot}}{f_{K^*}^{\Vert}}\frac{m_{s}+m_{q}}{m_{K^*}}.
\label{def:lambda}
\end{align}
For the  DA parameters we use the numerical update from Ref.\cite{Ball:2007zt} 
\begin{equation}
f_{K^*}^{\Vert} =220\text{ MeV},~\ \, f_{K^*}^{\bot}(1\text{ GeV})=185 \text{ MeV}, \, \, m_{K^{*0}}=896 \text{ MeV},
\end{equation}%
\begin{equation}
a_{1K^*}^{\Vert}(1\text{ GeV})=-0.03\pm0.02,~\ \ a_{1K^*}^{\bot}(1\text{ GeV})=-0.04\pm0.03,
\end{equation}%
\begin{equation}
a_{2K^*}^{\Vert}(1\text{ GeV})=0.11\pm0.09,~\ \ a_{2K^*}^{\bot}(1\text{ GeV})=0.10\pm0.08.
\end{equation}%
Notice that we define $SU(3)$ breaking coefficients $a_{1K^*}^{\Vert,\bot}$ to be negative because $K^{*}$ state  includes   $s$-antiquark. 

\section{Appendix B.  Analytical results for the collinear  convolution integrals}  
Here we provide results for the   convolution  integrals which describe colour-singlet  amplitude $A_{2c}^{\perp(0)}$. These integrals  have been  computed using DAs  described in the previous section.  All the divergent integrals are computed using analytical regularisation prescription as described in the text. The singular terms (poles in $1/\varepsilon$) are subtracted. The factorisation scales are fixed to be equal  $m_c$.  The resulting expressions have  subscript ``fin''.  For simplicity we do not write explicitly  the dependence on the factorisation scale in the parameters of DAs. 

 For the colour-singlet integrals in Eqs.(\ref{res:A0Jbot})   we obtain
 \begin{align}
&J_{c}^{(J=2)}[P_{2}V_{3}]_{\text{fin}}=
\\
&\left\{ \Delta\phi_{2P}^{\prime}(0)I[\bar{\Omega}]+\bar{\phi}_{2P}^{\prime}(0)I[\Delta\Omega]\right\}\frac12 \left(  1-\ln2\right)
+\left\{  \bar{\phi}_{2P}^{\prime}(0)\Delta\Omega
(1)+\Delta\phi_{2P}^{\prime}(0)\bar{\Omega}(1)\right\}
\frac12 \left(  1-\frac{\pi^{2}}{12}\right) 
\nonumber \\ & 
+\frac{27}{16}b_{1}\left\{8+3\pi^{2}-16\ln2+a_{2}^{\Vert}\left(  93-13\pi^{2}/4-16\ln2\right)
\right\} 
\nonumber \\
& -\frac{27}{16}b_{1}\lambda_{s}^{+}\left\{  ~6\zeta(3)-16+32\ln2+\pi^{2}%
(4\ln2-6) 
 +a_{2V}^{\bot}\left[  36\zeta(3)-516+352\ln2+\pi^{2}(24\ln
2-41)\right]  ~\right\}  
\nonumber \\ 
&+\frac{9}{16}a_{1V}^{\Vert}\left(  16\ln2-40-\pi^{2}+\frac{3}{2}b_{2}\left\{
260-49\pi^{2}+64\ln2\right\}  \right)
\nonumber \\ 
&+\frac{9}{16}\lambda_{s}^{-}\left(  -32\ln2+4\pi^{2}\ln2+6\zeta(3)+3b_{2}%
\left\{  -64\ln2+\pi^{2}(-15+\frac{32}{3}\ln2)+12\left(  10+\zeta(3)\right)
\right\}  \right)
\nonumber \\ 
& +\frac{9}{16}\lambda_{s}^{-}a_{2V}^{\bot}\left\{ 36\zeta(3)+400+2\pi
^{2}(5+16\ln2)-352\ln2
+3b_{2}\left[  72\zeta(3)-580-704\ln
2+\pi^{2}(155+48\ln2)\right] \right\}
\nonumber \\ 
& +\frac{27}{32}a_{1V}^{\bot} \lambda_{s}^{+}\left\{ -12\zeta(3)-80+96\ln
2-2\pi^{2}(1+4\ln2)
+b_{2}\left[  -72\zeta(3)+60+576\ln2-3\pi^{2}(19+16\ln2)\right] \right\}  .
\nonumber
\end{align}
\begin{align}
J_{c}^{(J=2)}[P_{3}V_{2}]_{\text{fin}} =& \bar{\phi}_{2V}^{\bot\prime}(0)\rho_{-}^{K}\left(-\frac32\right)\left( 15b_2 +(1+6b_2)\frac{\pi^{2}}{12}%
-(1+21b_2)\ln2\right)  +\Delta\phi_{2V}^{\bot\prime}(1)\left(1-\ln2\right)  
\nonumber \\
&-9 a^\perp_{1V}(2\ln2-4)-\frac92\rho_{-}^{K}(\pi^{2}+4\ln2)(1+6a^\perp_{2V})
\nonumber \\
&+\frac{27}4\rho_{-}^{K}b_{2}(1+6a^\perp_{2V})(20+3\pi^{2}-28\ln2).
\end{align}
The  integrals for the vector state $J=1$  can be obtained using  Eqs.(\ref{res:sumJcP2V3}) and (\ref{res:sumJcP3V2}). 

\end{document}